\documentclass[]{aastex631}
\shorttitle{XRT~210423 -- Host candidates}
\shortauthors{Eappachen et al.}
\graphicspath{{./}{figures/}}

\def\xrt{XRT~210423}
\def\cdf{CDF$-$S~XT2}
\def\lum{erg~s$^{-1}$}
\def\arcmin{\hbox{$^\prime$}}
\def\arcsec{\hbox{$^{\prime\prime}$}}
\def\approxlt{\ifmmode \rlap{$<$}{}_{{}_{{}_{\textstyle\sim}}} \else%
$\rlap{$<$}{}_{{}_{{}_{\textstyle\sim}}}$\fi}
\def\msun{M_\odot}

\begin{document}

\title{The Fast X-ray Transient XRT~210423 and its Host Galaxy}

\author[0000-0001-7841-0294]{D.~Eappachen}
\affiliation{SRON, Netherlands Institute for Space Research, Niels Bohrweg 4, 2333 CA, Leiden, The Netherlands}
\affiliation{Department of Astrophysics/IMAPP, Radboud University Nijmegen, P.O. Box 9010, 6500 GL, Nijmegen, The Netherlands}

\author[0000-0001-5679-0695]{P.G.~Jonker}
\affiliation{Department of Astrophysics/IMAPP, Radboud University Nijmegen, P.O. Box 9010, 6500 GL, Nijmegen, The Netherlands}
\affiliation{SRON, Netherlands Institute for Space Research, Niels Bohrweg 4, 2333 CA, Leiden, The Netherlands}

\author{A.J.~Levan}
\affiliation{Department of Astrophysics/IMAPP, Radboud University Nijmegen, P.O. Box 9010, 6500 GL, Nijmegen, The Netherlands}

\author[0000-0001-8602-4641]{J.~Quirola-V\'asquez\ }
\affiliation{Instituto de Astrof\'isica, Pontificia Universidad Cat\'olica de Chile, Casilla 306, Santiago 22, Chile}
\affiliation{Millennium Institute of Astrophysics (MAS), Nuncio Monse$\tilde{n}$or S\'otero Sanz 100, Providencia, Santiago, Chile}
\affiliation{Observatorio Astron\'omico de Quito, Escuela Polit\'ecnica Nacional, 170136, Quito, Ecuador}

\author[0000-0002-5297-2683]{M.A.P.~Torres}
\affiliation{Instituto de Astrof\'{i}sica de Canarias, E-38205 La Laguna, S/C de Tenerife, Spain}
\affiliation{Departamento de Astrof\'isica, Univ. de La Laguna, E-38206 La Laguna, Tenerife, Spain}

\author[0000-0002-8686-8737]{F.E.~Bauer}
\affiliation{Instituto de Astrof\'isica, Pontificia Universidad Cat\'olica de Chile, Casilla 306, Santiago 22, Chile}
\affiliation{Millennium Institute of Astrophysics (MAS), Nuncio Monse$\tilde{n}$or S\'otero Sanz 100, Providencia, Santiago, Chile}
\affiliation{Space Science Institute, 4750 Walnut Street, Suite 205, Boulder, Colorado 80301, USA}

\author{V.S.~Dhillon}
\affiliation{Department of Physics \& Astronomy, University of Sheffield, Sheffield S3 7RH, UK}
\affiliation{Instituto de Astrof\'{i}sica de Canarias, E-38205 La Laguna, S/C de Tenerife, Spain}

\author{T.~Marsh}
\affiliation{Department of Physics, Gibbet Hill Road, University of Warwick, Coventry, CV4 7AL, UK}

\author{S.P.~Littlefair}
\affiliation{Department of Physics \& Astronomy, University of Sheffield, Sheffield S3 7RH, UK}

\author[0000-0003-3193-4714]{M.E.~Ravasio}
\affiliation{Department of Astrophysics/IMAPP, Radboud University Nijmegen, P.O. Box 9010, 6500 GL, Nijmegen, The Netherlands}

\author[0000-0003-2191-1674]{M.~Fraser}
\affiliation{School of Physics, O’Brien Centre for Science North, University College Dublin, Belfield, Dublin 4, Ireland}




\begin{abstract}

Fast X-ray Transients (FXTs) are X-ray flares with a duration ranging from a few hundred seconds to a few hours. Possible origins include the tidal disruption of a white dwarf by an intermediate-mass black hole, a supernova shock breakout, and a binary neutron star merger. We present the X-ray light curve and spectrum, and deep optical imaging of the FXT XRT~210423, which  has been suggested to be powered by a magnetar produced in  a binary neutron star merger. Our \textit{Very Large Telescope} and \textit{Gran Telescopio Canarias} (GTC) observations began on May 6, 2021, thirteen days after the onset of the flare. No transient optical counterpart\footnote{We use the word "counterpart" for any transient light in a waveband other than the original X-ray detection wave band, whereas the word "host" refers to the host galaxy.} is found in the 1\arcsec{} (3$\sigma$) X-ray uncertainty region of the source to a depth $g_{s}$=27.0 AB mag. A candidate host lies within the 1\arcsec{} X-ray uncertainty region with a magnitude of 25.9 $\pm$ 0.1 in the GTC/HiPERCAM $g_s$-filter. Due to its faintness, it was not detected in other bands, precluding a photometric redshift determination.We detect two additional candidate host galaxies; one with $z_{\rm spec}=1.5082 \pm 0.0001$ and an offset of 4.2$\pm$1\arcsec{} (37$\pm$9 kpc) from the FXT  and another one with $z_{\rm phot}=1.04^{+0.22}_{-0.14}$, at an offset of 3.6$\pm$1\arcsec{} (30$\pm$8 kpc). Based on the properties of all the prospective hosts we favour a binary neutron star merger, as previously suggested in the literature, as explanation for XRT~210423.
\end{abstract}



\section{Introduction} \label{sec:intro}
Fast X-ray transients (FXTs) have been discovered through systematic searches in {\it Chandra}, XMM-{\it Newton}, and {\it eROSITA} data by \citet{2013jonker}; \citet{glennie}; \citet{Irwin}; \citet{Bauer}; \citet{Xue2019}; \citet{Alp2020}; \citet{Novara2020}; \citet{2020ATel13416....1W}; \citet{Quirola2022}; \citet{2022ApJ...927..211L}. FXTs manifest as short flashes of X-rays, with a duration ranging from a few minutes to hours. Several different physical mechanisms have been suggested for the origin of FXTs. The leading theories explaining FXTs include events that are related to strong gravitational wave sources such as a binary neutron star merger (BNS; \citealt{2006Sci...311.1127D}; \citealt{2008MNRAS.385.1455M}) and a white dwarf (WD) disruption by an intermediate-mass black hole (IMBH; \citealt{2020SSRv..216...39M}), as well as supernovae shock-breakout (SBO) where those involving relatively compact progenitors are predicted to be detectable in soft X-rays (\citealt{2017hsn..book..967W}).

Some neutron star mergers are thought to leave behind a rapidly rotating neutron star with a strong magnetic field (i.e., a magnetar) which could produce an X-ray transient (\citealt{2006Sci...311.1127D}; \citealt{2008MNRAS.385.1455M}; \citealt{2013ApJ...763L..22Z}; \citealt{2017ApJ...835....7S}). \citet{Xue2019} favour this scenario for XRT 150321/\cdf. The apparent plateau in its X-ray light curve comparable to those seen in a subset of short gamma-ray bursts (SGRBs; \citealt{2013MNRAS.430.1061R}), has been used to argue for a BNS origin.


Tidal disruption events (TDEs) involving  a massive BH and a main sequence star have timescales ranging from hours to weeks (\citealt{2021SSRv..217...18S}). The relatively low mass of an IMBH compared to that of a super-massive black hole coupled with the compactness of a WD compared to that of a main sequence star is predicted to lead to a fast X-ray flash (\citealt{2009ApJ...695..404R}; \citealt{2020SSRv..216...39M}). Some of the FXTs have been interpreted as a WD IMBH TDEs (e.g., XRT~000519; \citealt{2013jonker}; \citealt{2019ApJ...884L..34P}). Precursor flares are found in XRT~000519, and their timescale can be explained by the expected orbital time scale of a WD in an eccentric orbit around an IMBH (\citealt{2013jonker}; \citealt{2020SSRv..216...39M}).

The radiation--mediated shock from a supernova (SN) explosion crossing the star's surface releases a so-called SBO (\citealt{2017hsn..book..967W}). Depending on the radius of the exploding star, the SBO could initially appear as an X-ray flash and then evolve into an UV and optical signal. The best example of a SBO is XRT~080109 (\citealt{2008Natur.453..469S}) in the galaxy NGC~2770, where the FXT was associated with a bright Type Ibc SN,  SN~2008D (\citealt{2008Sci...321.1185M}; \citealt{2009ApJ...702..226M}).

About 30 FXTs have been identified to date, both serendipitously and through careful searches in archival data. Counterpart searches have been done for some FXTs. For instance, in  the case of CDF-S~XT1, \citet{Bauer} did not find a counterpart down to deep optical limits from Very Large Telescope (VLT) imaging beginning 80 minutes after the onset of the X-ray flare (R$>$ 25.7 mag; \citealt{Bauer}). Consequently, deep host searches have been undertaken from both space and ground for other FXTs. For instance, deep VLT and Gran Telescopio Canarias (GTC) observations of XRT~000519 (\citealt{2013jonker}) were reported by \citet{2022MNRAS.514..302E}. They report the detection of a candidate host galaxy to the northwest side of the FXT position in the $g$-band image at $m_{g}$=26.29$\pm$0.09 AB mag. An SBO scenario is firmly ruled out if the FXT is indeed associated with this candidate host galaxy. If XRT~000519 is at the distance of  M86, it is unlikely that an associated optical SN was missed, while if at a larger redshift, the peak X-ray luminosity would be so large it would rule out an SN SBO origin. Deep optical observations have constrained the hosts of two other FXTs: CDF-XT1 (\citealt{Bauer}) lies at a 0.13$^{\prime\prime}$ offset from a $m_{r}$=27.5 dwarf galaxy with z$_{\rm phot}$=0.4-3.2 and, \cdf{} (\citealt{Xue2019}) lies at a projected distance of 3.3 $\pm$ 1.9 kpc from a $z_{\rm spec}$=0.74 star-forming galaxy. For XRT~170901 \citet{2022ApJ...927..211L} finds an elongated, possibly late-type, star-forming galaxy within the 2$\sigma$ X-ray uncertainty region while for XRT~030511 (\citealt{2022ApJ...927..211L}), no host galaxy was found.

In this paper, we present the X-ray light curve and spectrum of the FXT \xrt{}, our search for its optical counterpart and host galaxy, and the analysis of the potential host galaxies we identified. Throughout the paper, we use the concordance $\Lambda$-CDM cosmology, with Hubble constant  H$_0$= 67.4$\pm$0.5 km/s/Mpc and matter density parameter $\Omega_m$=0.315$\pm$0.007 (\citealt{2020A&A...641A...6P}). In \S~\ref{sec:obs}, we describe the observations and analysis, we present the results in \S~\ref{sec:results}, which we discuss in \S~\ref{sec:discuss}, and draw conclusions in \S~\ref{sec:conclusion}.

\section{OBSERVATIONS AND ANALYSIS} \label{sec:obs}

\subsection{\textit{Chandra} X-ray data}

The FXT \xrt{} was serendipitously discovered on 2021 April 23 (\textit{Chandra} Observation ID: ObsID 24604), in the direction of Abell~1795 in \textit{Chandra} archival data and reported by \citet{2021ATel14599....1L}. The transient position was covered by the I3 CCD of the ACIS-I array of CCD detectors. We reprocessed and analysed the data with the {\sc CIAO 4.14} software developed by the Chandra X-Ray Center employing CALDB version 4.9.1. To identify the X-ray source, determine its position and its associated uncertainty, we utilize the {\sc CIAO} source detection tool \texttt{wavdetect} 
\citep{Freeman2002} using a series of ``Mexican Hat'' wavelet functions to account for the varying PSF size across the detector. To improve the \emph{Chandra} astrometric 
solution, we cross-match the positions of fifteen X-ray sources with those of their prospective optical counterparts that either come from the \emph{Gaia} Early Data Release 3 (\emph{Gaia}-EDR3; \citealt{Gaia_DR3_2020}) or the Sloan Digital Sky Survey Data Release 16 (SDSS--DR16;\citealt{Ahumada2019})
catalogues. For this we use the \texttt{wcs\_match} script in {\sc CIAO}.
The average residual is $0.82{\pm}0.10$~arcsec.

We extracted the source X-ray spectrum with the \texttt{specextract} package in {\sc CIAO}. For this, we included the X-ray counts within a circular region centered at the X-ray position with a radius of 9.2\arcsec{} (corresponding to an encircled energy fraction of ${\approx}$98\% given its off-axis angle of 7.5\arcmin{}; \citealt{Yang2019}). For the background regions, we considered two adjacent rectangular regions (centered on R.A.$_{\rm J2000.0}{=}$13$^{\rm h}$48$^{\rm m}$59$^{\rm s}$.27, Dec$_{\rm J2000.0}{=}$26$^\circ$39$\arcmin$33$\arcsec$.9 and R.A.$_{\rm J2000.0}{=}$13$^{\rm h}$48$^{\rm m}$53$^{\rm s}$.55, Dec$_{\rm J2000.0}{=}$26$^\circ$39$\arcmin$55$\arcsec$.3, size of 52\arcsec{}${\times}$17\arcsec{}, and a position angle of the long side of $15^\circ$ regarding the east-west axis) to cover the gradient emission of the galaxy cluster LEDA 94646, which is the main source of background photons. This procedure yields a count rate of ${\approx}9.7{\times}10^{-3}$~cts~s$^{-1}$ for \xrt{}. Using the Bayesian X-ray Astronomy package \citep[BXA;][]{Buchner2014}, within the fitting environment of \texttt{XSPEC} version 12.10.1f \citep{Arnaud1996} and Cash statistics \citep{Cash1979}, we fitted the spectra of \xrt{}. We used an absorbed power-law model (\texttt{phabs*zphabs*pow} model in \texttt{XSPEC}) to describe the data, where \texttt{phabs} and \texttt{zphabs} describe the Galactic and intrinsic absorption, respectively. For the extraction of the X-ray spectral parameters, we fixed the Galactic hydrogen column density ($N_{H,\rm Gal}$) to $3.0\times$10$^{20}$~cm$^{-2}$ \citep[taken from][]{Kalberla2005, Kalberla2015}, while the intrinsic hydrogen column density ($N_H$) is a free parameter in our fit. In addition, we calculate the hardness ratio (see Sect.~\ref{sec:X_ray}), which we define as HR=($H$-$S$)($H$+$S$)$^{-1}$, where $H$ and $S$ are the count rates in the 2.0--7.0~keV and 0.5–2.0~keV bands, respectively, and we derive their errors based on the Bayesian code \texttt{BEHR} \citep{Park2006}.

\subsection{Optical data}
We observed the field containing the FXT \xrt{} with the FOcal Reducer/low dispersion Spectrograph (FORS2; \citealt{1998Msngr..94....1A}) mounted on the 8.2~m Very Large Telescope (VLT) of the European Southern Observatory (ESO).  
FORS2 on VLT  has a field of view 6.8\arcmin{} $\times$ 6.8 \arcmin{} and  provides a pixel scale of  0.25\arcsec{} with 2 $\times$ 2 binning. A total of 7 $\times$ 180~s images were taken on 2021 May 6 in the Johnson-Cousins $R$-band filter with the two 2k×4k MIT CCDs.

The average seeing in the images was around 1.8\arcsec. The ESO reflex data reduction pipeline (\citealt{2013A&A...559A..96F}) was used for bias and flat field correction. Next, we employed $\textsc{L.A.Cosmic}$ to remove charge caused by cosmic ray hits from the images (\citealt{Lacos2001PASP..113.1420V}) and stacked them using the  {\textsc{IRAF IMCOMBINE}} task. We applied an astrometric correction to the stacked image using astrometry.net\footnote{https://astrometry.net/}(\citealt{2010AJ....139.1782L}).

On May 13 and June 10, 2021, we also obtained simultaneous $u_s$, $g_s$, $r_s$, $i_s$, and $z_s$-band images with the HiPERCAM instrument mounted on the 10.4~m Gran Telescopio Canarias (GTC) at the  Roque de los Muchachos Observatory (La Palma, Spain).  Seventeen frames with an exposure time of 180~s each 
were obtained. HiPERCAM with 2 $\times$ 2 binning provides a plate scale of 0.16\arcsec~per pixel and a field of view of 2.8\arcmin{} $\times$ 1.4\arcmin{} on GTC (\citealt{hipercam2021MNRAS.507..350D}). The average seeing in the images was around 1.4\arcsec{} and 0.9\arcsec{} for the first and second epoch, respectively.

The HiPERCAM data reduction steps were performed using a dedicated data reduction pipeline, including bias subtraction, flat field and fringe correction.\footnote{https://cygnus.astro.warwick.ac.uk/phsaap/HiPERCAM/docs/html/}  For both epochs, the 17 images were averaged to create a single deep image using the {\textsc{IRAF IMCOMBINE}} task. We scaled each image to a common level (scale='mode') before using reject='avsigclip' to remove charge caused by cosmic rays. 

For each of the combined  $u_s$, $g_s$, $r_s$, $i_s$, and $z_s$-band HiPERCAM images, we refined the world coordinate information that was based on the
telescope pointing using the known astrometric positions of 8 stars
from the Pan-STARRS catalogue (\citealt{2016arXiv161205560C}). We used the centroid algorithm in
\textsc{IRAF PHOT} and \textsc{ IRAF CCMAP} to determine the world coordinate solution and applied the astrometric corrections using \textsc{IRAF CCSETWCS}.

\begin{figure*} 
    \centering
	\includegraphics[scale=0.60]{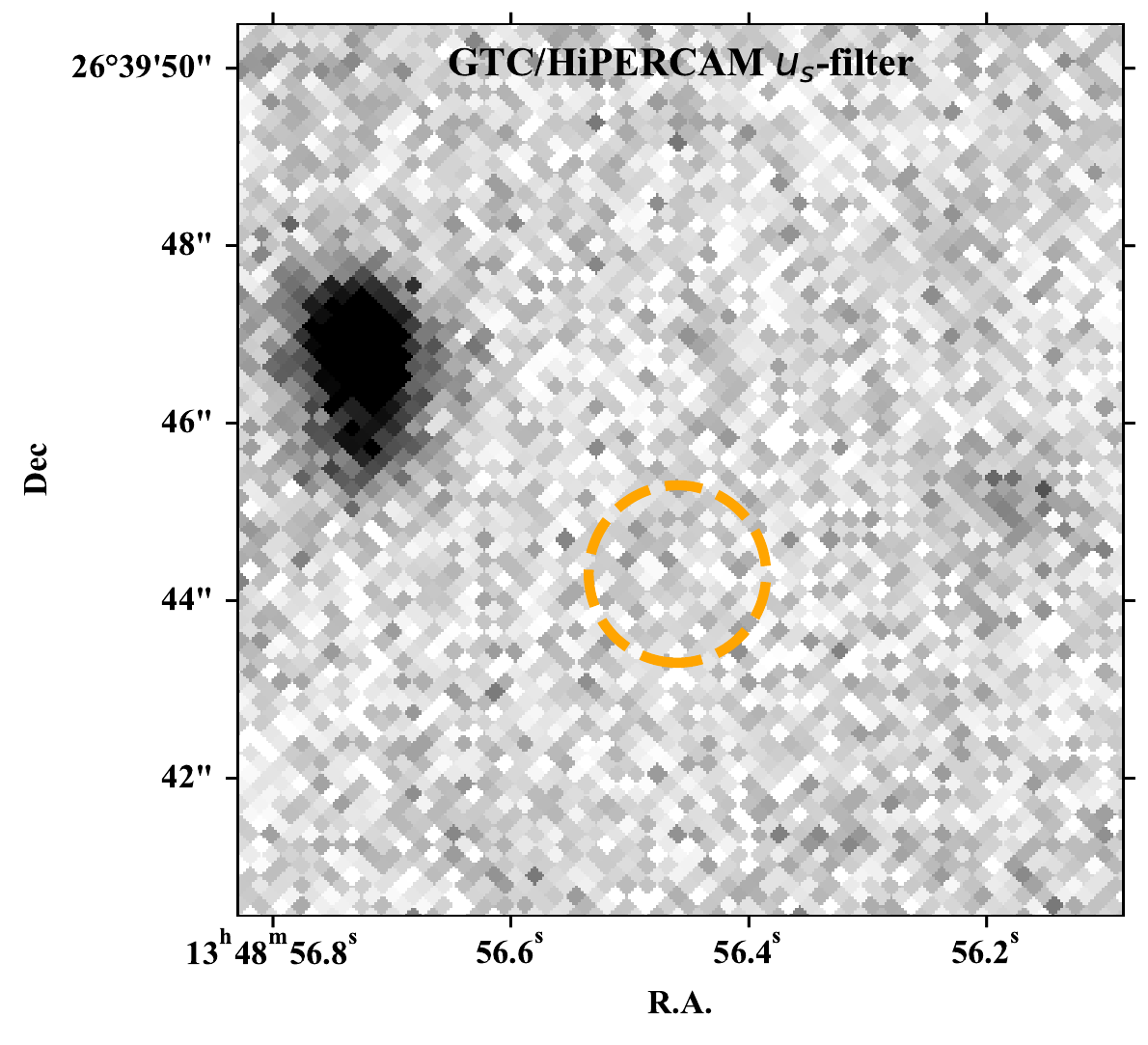}
    \includegraphics[scale=0.60]{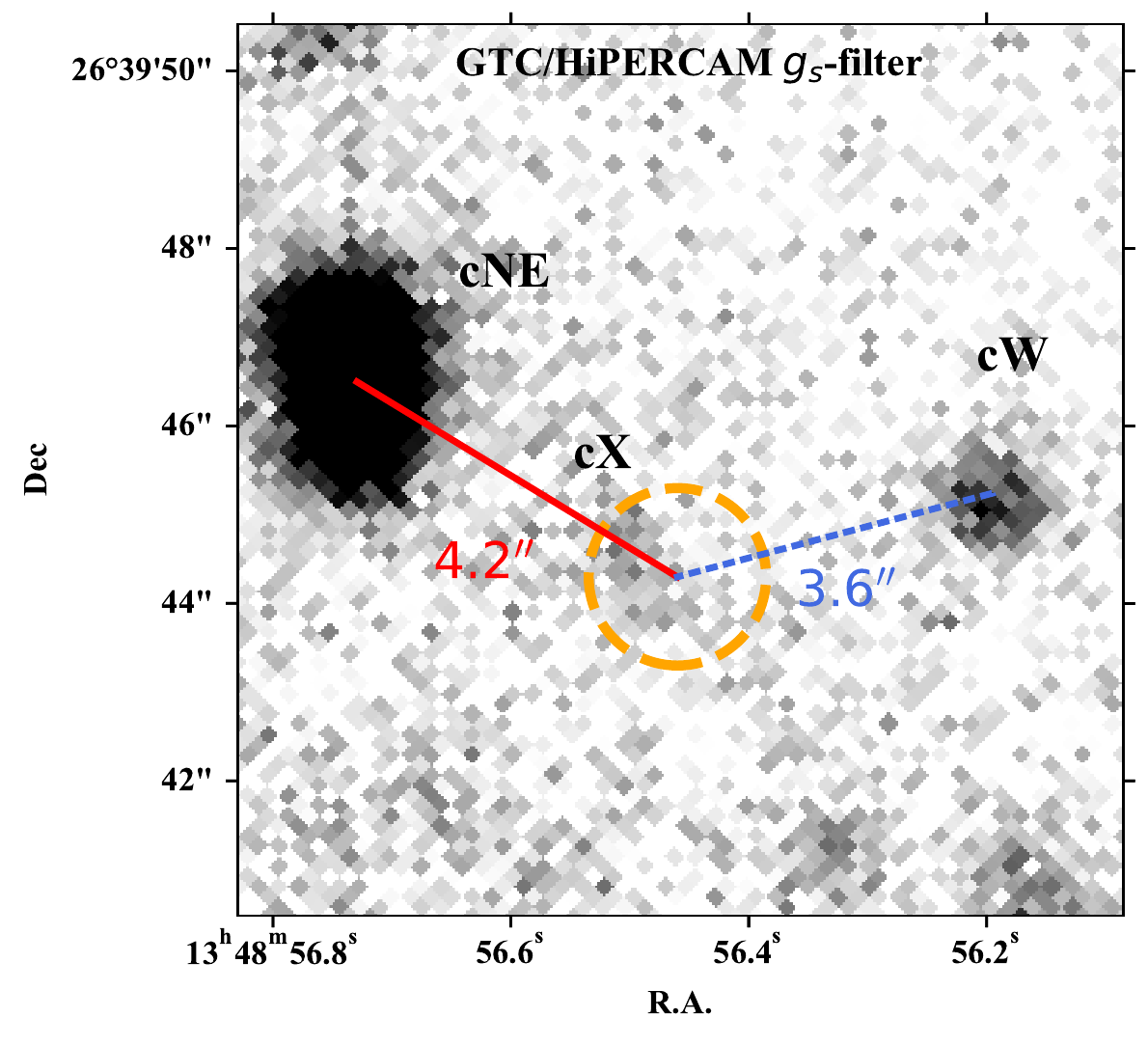}
    \includegraphics[scale=0.60]{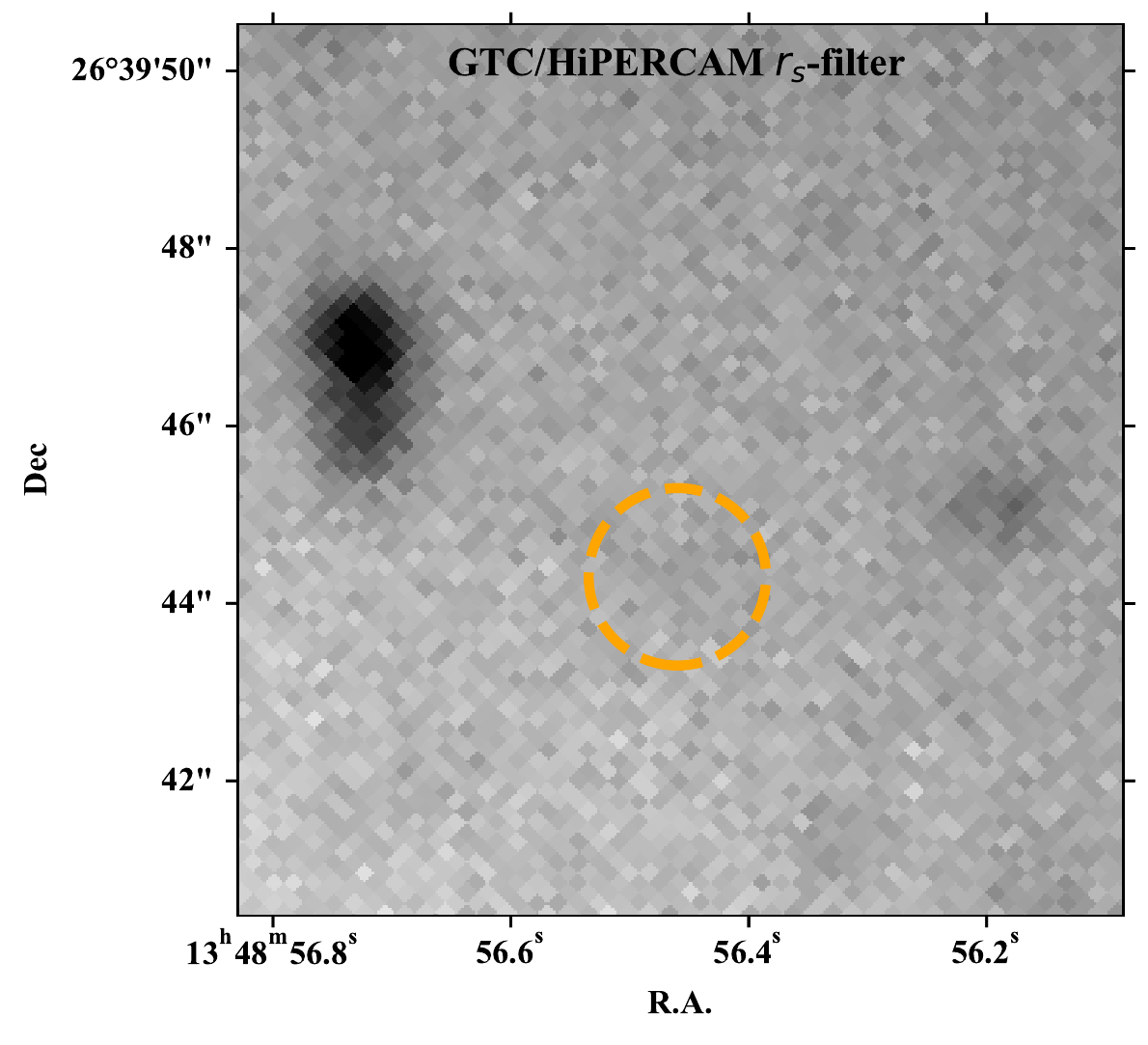}
    \includegraphics[scale=0.60]{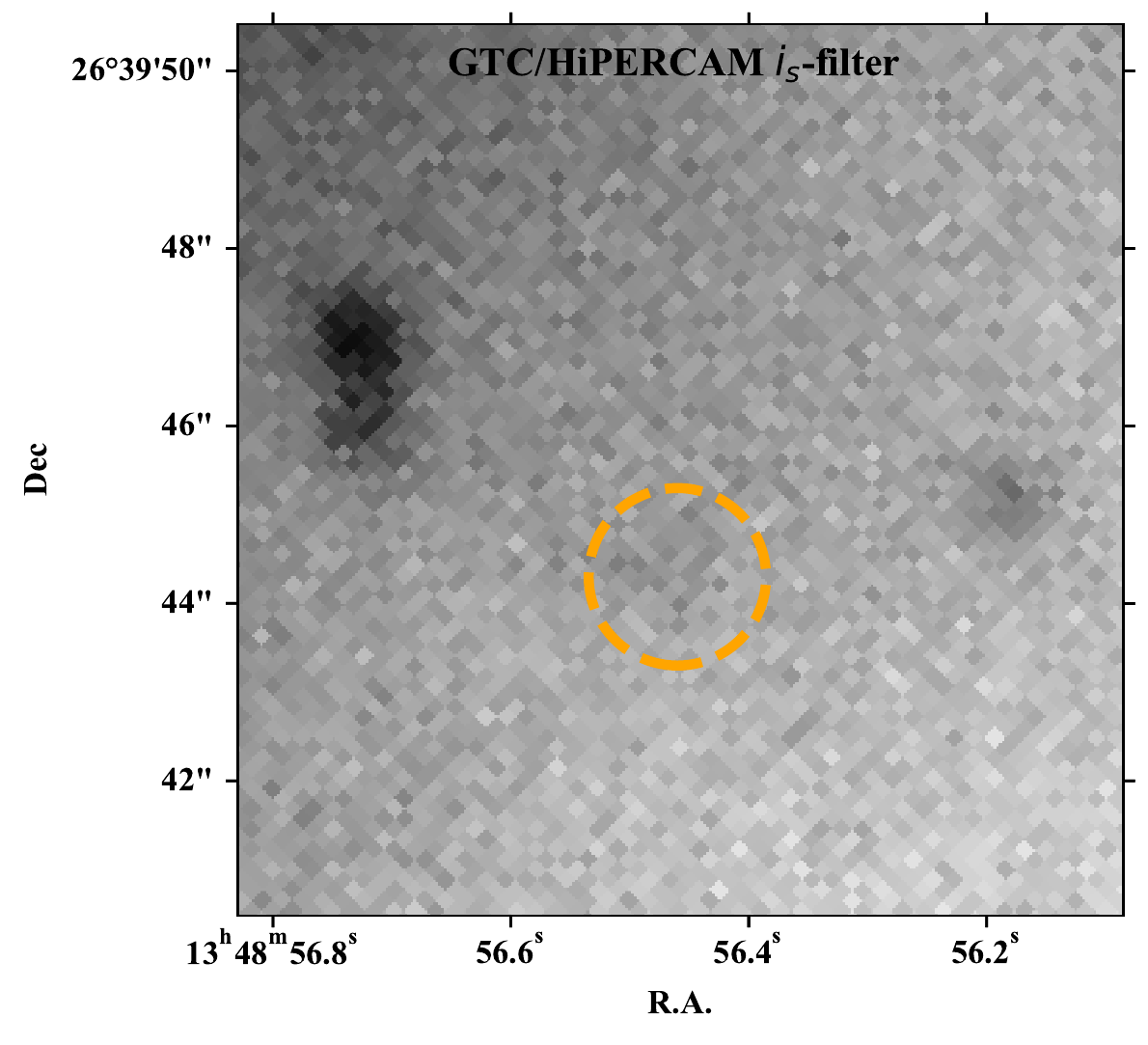}
    \includegraphics[scale=0.60]{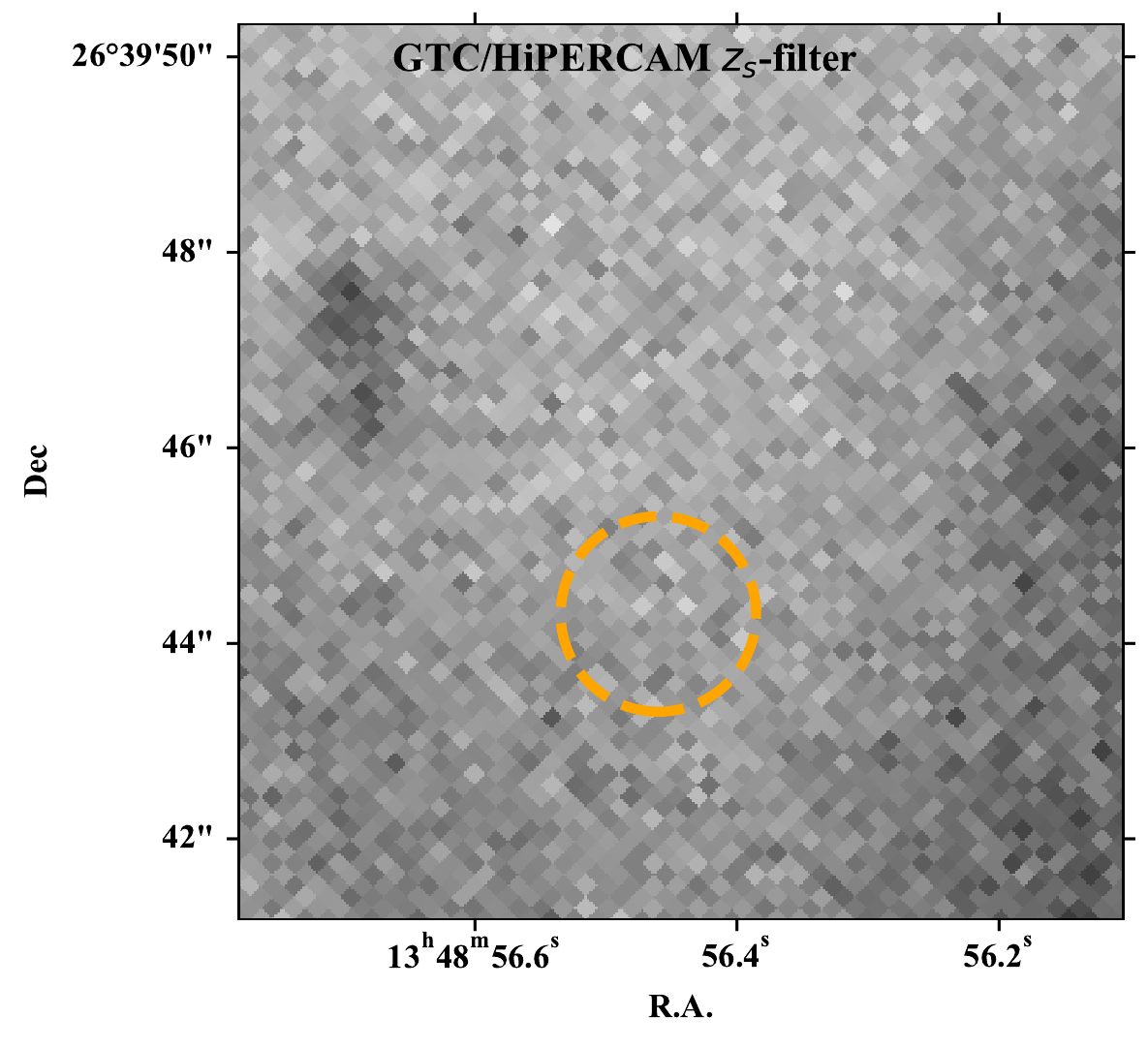}
    \includegraphics[scale=0.60]{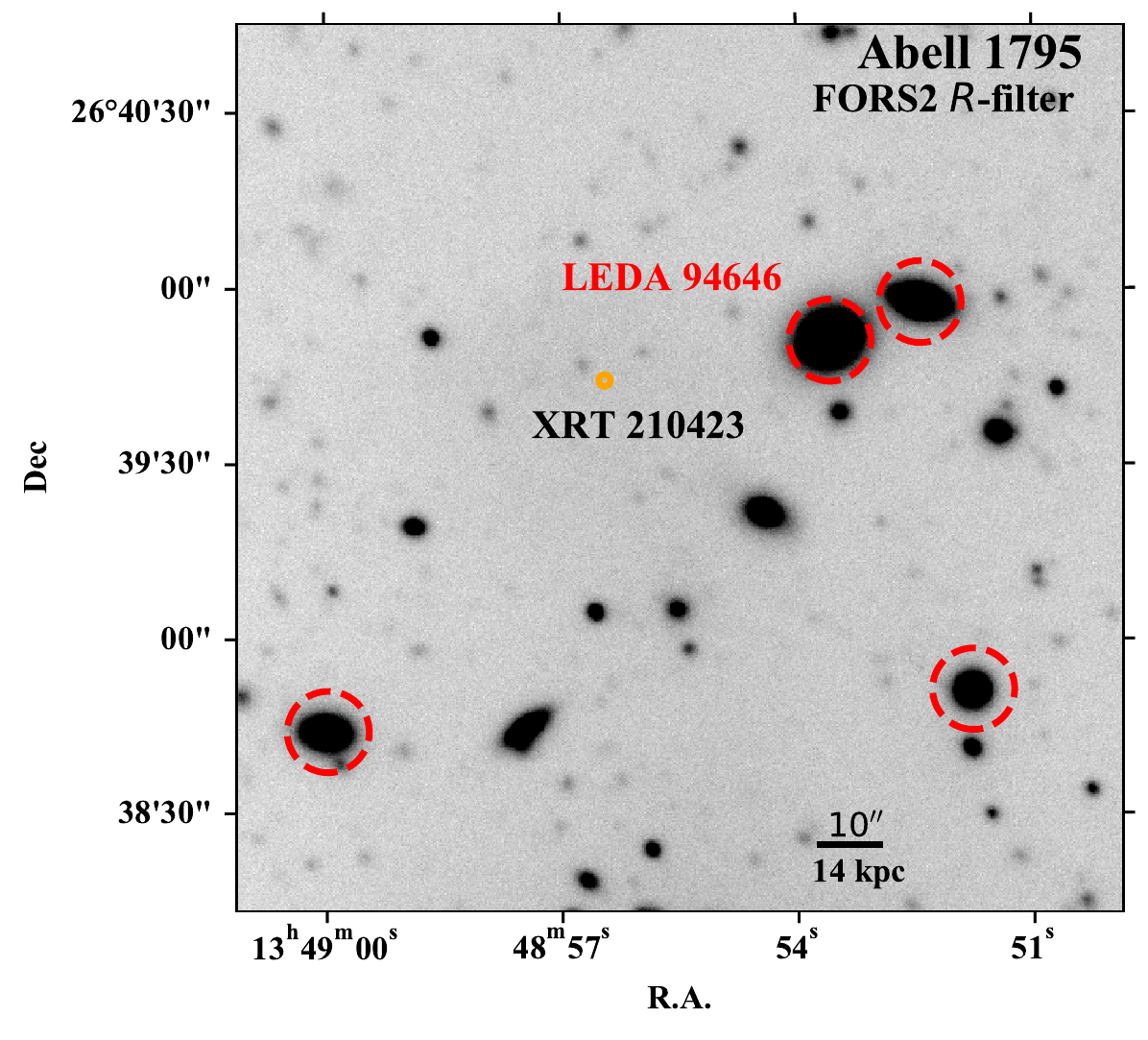}    \caption{The second epoch GTC/HiPERCAM, and the VLT/FORS2 images of the field of \xrt{}. From left to right and top to bottom: the GTC $u_s$, $g_s$, $r_s$, $i_s$, $z_s$ and the VLT $R$-band images. The dashed orange circle shows the FXT 3$\sigma$ X-ray uncertainty position. The \textit{top right} panel with the GTC/HiPERCAM $g_s$-filter image shows the positional offset between the center of the \xrt{} X-ray uncertainty region and the host candidates cNE (indicated by the red line) and cW (the dashed blue line). In addition, the faint candidate host cX can (just) be seen within the $3\sigma$ FXT position. The \textit{bottom right} panel shows the 
    position of \xrt{} with respect to the Abell~1795 cluster in  VLT/FORS2 $R$-filter. The FXT position is marked with an orange circle, while several members of the cluster are indicated with red dashed circles. }
    
    \label{fig:210423}
\end{figure*}

We used $\textsc{SExtractor}$ (\citealt{1996A&AS..117..393B}) to extract the R.A.~and Dec, magnitude, and magnitude error from all the objects detected in the stacked image. Photometric calibration of the image was done using stars from the Pan-STARRS DR2 catalogue data for $g_s$, $r_s$, $i_s$, $z_s$ and    $R$ filters (\citealt{2016arXiv161205560C}). For the zero point value of the $u_s$-filter, we used SDSS catalogue data.
For the FORS2 image we used the transformation equation of \citet{Lupton:2005} to convert the Pan-STARRS magnitudes given in the Sloan $r$- and $i$ bands to a Johnson-Cousins $R$-band magnitude. The resultant GTC/HiPERCAM $u_s$, $g_s$, $r_s$, $i_s$, $z_s$, and the VLT/FORS2 $R$-filter images of the field of \xrt{} are shown in Figure \ref{fig:210423}.

We obtained a spectrum of the galaxy SDSS~J134856.75+263946.7, which lies  $\sim$4.2$^{\prime\prime}$ from the position of the FXT \xrt{} on May 16, 2021 using the X-SHOOTER instrument mounted on the UT3 VLT (\citealt{2011A&A...536A.105V}). X-SHOOTER is a multi-wavelength (300-2500 nm), medium--resolution spectrograph mounted at the Cassegrain focus. Using slit widths of 1.3$^{\prime\prime}$/1.2$^{\prime\prime}$/1.2$^{\prime\prime}$, we obtained exposures of 2$\times$ 735 sec, 2$\times$680 sec, and 4$\times$300 sec in the UVB, VIS, and NIR arm of the spectrograph, respectively. For data reduction, we used the ESO Reflex pipeline (\citealt{2013A&A...559A..96F}).

\begin{figure*} 
    \centering
	\includegraphics[scale=0.55]{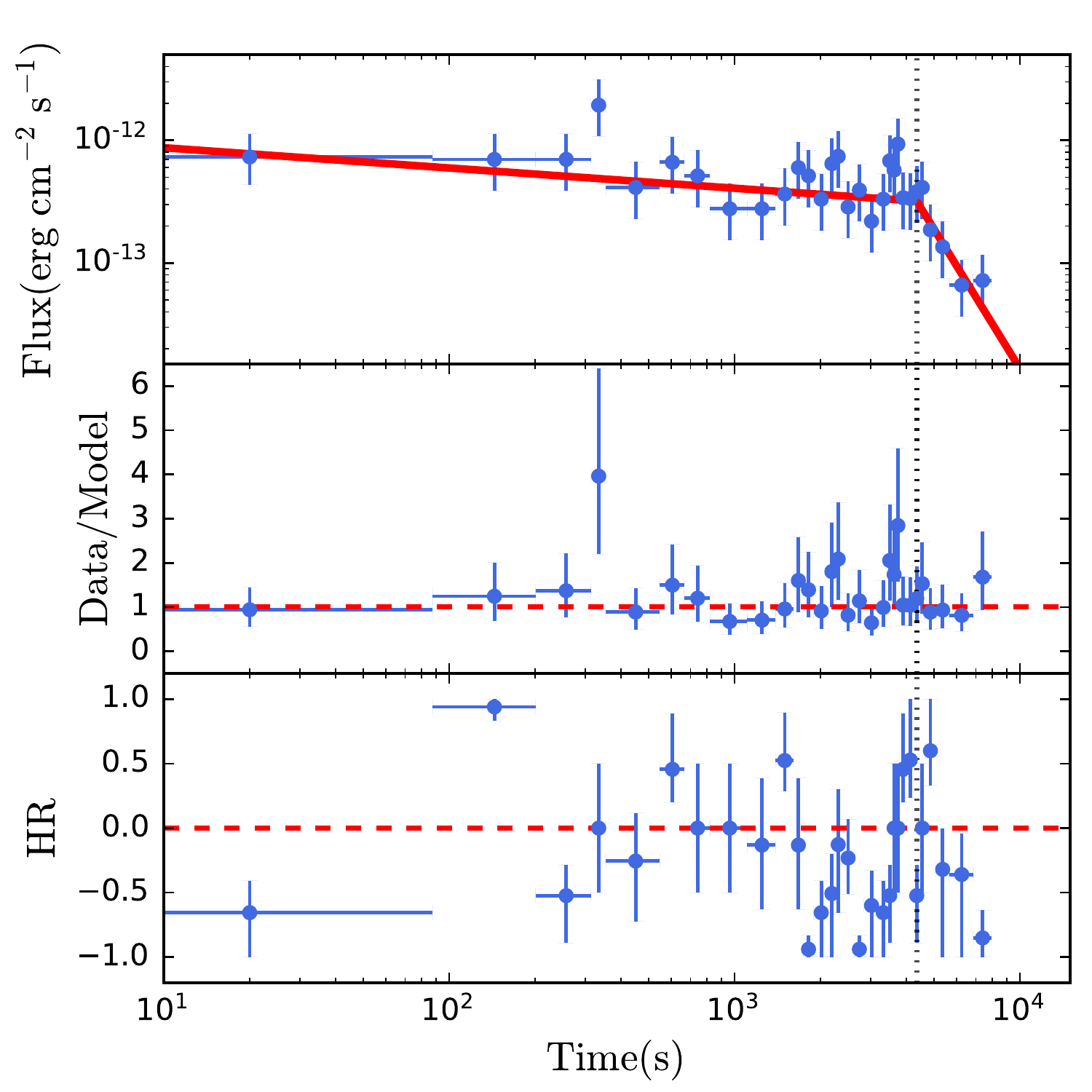}
	\includegraphics[scale=0.55]{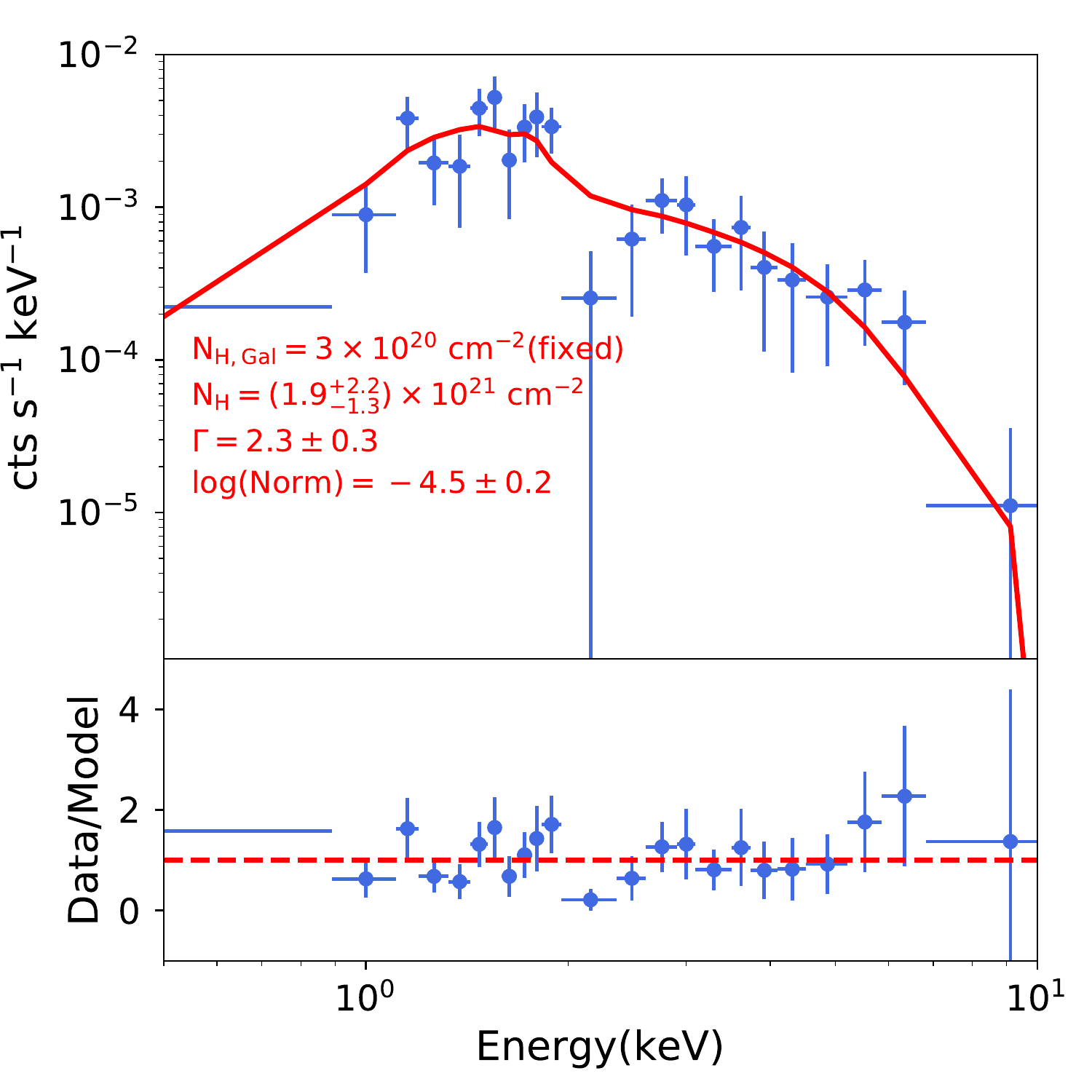}
    \caption{X-ray flux and spectral evolution of \xrt{}. 
    \textit{Left top panel:} 0.5-7 keV light curve (blue points and 1$\sigma$ errors) with the best-fit broken power-law model (\textit{red solid line}).
    \textit{Left middle panel:} The ratio between the data and the best-fit model light curve.
    \textit{Left bottom panel:} The hardness-ratio (HR) evolution. In all three panels the vertical dashed line ($t_{B} = 4350$~s) indicates the best-fit break time. We compute the light-curve zero point ($T_0{=}0$~sec), dividing the light curves in bins of $\Delta t{=}$100 and 10~seconds, and compute the chance probability that the photons per bin come from the background ($P_{\rm bkg}$) using the Poisson probability and the background rate. We found that the bins after $T_0$ have a $P_{\rm bkg}{\lesssim}$0.01, while $P_{\rm bkg}$ immediately before $T_0$ is higher $P_{\rm bkg}{\gtrsim}$0.1--0.2.
    \textit{Right top panel:} Average X-ray spectrum (blue points and 1$\sigma$ errors) and the best-fit absorbed power-law  model (red solid line). The text in the figure gives the parameters of the best-fit model (see main text for details). \textit{Right bottom panel:} The ratio between the data and the best-fit model. }
    \label{fig:210423_LC_SPEC}
\end{figure*}

\section{RESULTS} \label{sec:results}

\subsection{The X-ray light curve and spectrum}\label{sec:X_ray}
The \textit{Chandra} X-ray position of \xrt{} is R.A.$_{\rm J2000.0}{=}$13$^{\rm h}$48$^{\rm m}$56$^{\rm s}$.46 and Dec$_{\rm J2000.0}{=}$26$^\circ$39$\arcmin$44$\arcsec$.3 (with a 3$\sigma$ positional uncertainty of 1$\arcsec$; with an X-ray localization uncertainty of 0.31\arcsec from \texttt{wavdetect} and an uncertainty of 0.14\arcsec~associated with tying the source position to the International Celestial Reference System (ICRS), both at $1\sigma$ confidence level; \citealt{Rots2011}). Our resulting X-ray source position is  consistent with that of \citet{2021ATel14599....1L}. We do not detect the source in any other publicly available \textit{Chandra} observation covering the field of \xrt{}. The data spans a period from ${\approx}$13~years before the outburst to ${\approx}$40~days thereafter. Indeed, \textit{Chandra} revisited the \xrt{} field ${\approx}$40 days after the discovery observation of \xrt{} (ObsId~25049 DDT proposal, PI Dacheng Lin) with an exposure time of 59.3~ks. A 3-$\sigma$ upper limit of $F_X{\lesssim}$1.4$\times$10$^{-15}$~erg~cm$^{-2}$~s$^{-1}$ for the X-ray flux in the 0.5--7~keV band was obtained. Furthermore, the upper limit to any  persistent flux of \xrt{} is $F_X{\approxlt}10^{-15}$~erg~cm$^{-2}$~s$^{-1}$ in the 0.5--7~keV band at 3$\sigma$ confidence level. For this upper limit determination we considered all previous \emph{Chandra} observations of the field  yielding a total exposure time of ${\sim}2$~Ms (for stacking, we used the \texttt{merge\_obs} CIAO script), and the statistic developed by \citet{Kraft1991} at 3$\sigma$ confidence level.

Figure~\ref{fig:210423_LC_SPEC}, \textit{left panels}, shows the \textit{Chandra} 0.5–7~keV light curve of \xrt{}, and its best-fit (broken power-law) model. The light curve of \xrt{} is comprised of a net (background subtracted) number of  $116.1_{-11.8}^{+12.4}$ photons. The X-ray flash started at about 22:15:40 UT on April 23, 2021 ($T_0$) and lasted for ${\approx}$20--25~ks during a ${\approx}$26.4~ks \emph{Chandra} observation. The time span over which 5\% to 95\% of the total detected number of counts was registered, $T_{\rm 90}$, is  12.1$^{+4.0}_{-4.1}$~ks. The light curve is well fit by a broken power-law model (the chi-squared, the number of degrees-of-freedom and the Bayesian Information Criteria are 17.9, 26 and $-1.74$, respectively), with  slopes of $-0.17{\pm}0.06$ before the break at $t_{\rm break}{=}4.3{\pm}0.4$~ks and $-3.78{\pm}1.22$ after the break (see Fig.~\ref{fig:210423_LC_SPEC}). The pre-break average flux is  ${\rm F_X}{\approx}$7$\times$10$^{-13}$~erg~cm$^{-2}$~s$^{-1}$, while a minimum flux of ${\rm F_X}{\approx}$6--7$\times$10$^{-14}$~erg~cm$^{-2}$~s$^{-1}$ is measured in the last bin after the break.

The best-fitting absorbed power--law spectral model is shown in the \textit{right panel} of Fig~\ref{fig:210423_LC_SPEC}. The best-fit powerlaw has an index of $2.3{\pm}0.3$. The spectrum has an average hardness ratio of HR=$-0.2{\pm}0.1$. There is no evidence at the 3$\sigma$ level for spectral evolution over the flare (see Fig.~\ref{fig:210423_LC_SPEC}, \textit{bottom left panel}). In fact, fits to the spectra extracted before and after the break (dashed vertical line the \textit{left panel} of Fig.~\ref{fig:210423_LC_SPEC}) have consistent power-law spectral indices with values of $2.4{\pm}0.5$ and $2.1\pm0.7$, respectively. Here, the uncertainties are given at a 90\% confidence level.

\subsection{Candidate host identification and photometry} \label{sec:results_optical}
 We do not discuss the first epoch observation due to the relatively lower quality of data (seeing of $\sim$ 1.4\arcsec{}). In the second epoch  GTC observation on 2021 June 10, we detect a candidate host in the GTC/HiPERCAM $g_s$-filter
within the $\sim 1^{\prime\prime}$ circular error circle (3$\sigma$ confidence) of the \xrt{} at R.A. = $13^{\rm h}48^{\rm m}56.46^{\rm s}$, Dec =
$+26^{\rm\circ} 39^{\rm\prime}44.3^{\prime\prime}$ (hereafter called cX; see the GTC/HiPERCAM $g_s$-filter panel in Figure~\ref{fig:210423}). 
Using $\textsc{SExtractor}$, we obtain an aperture (15 pixels corresponding to 2.4$^{\prime\prime}$ in diameter) magnitude  of 25.9 $\pm$ 0.1 for this candidate host in the $g_s$ band.

We also measured the magnitude of other possible host candidates near the \xrt{} 3$\sigma$ uncertainty region. From the GTC/HiPERCAM and FORS2 images, we obtained the Kron magnitude for the candidate host that lies to the North-East (SDSS J134856.75+263946.7; hereafter called cNE) of the FXT position at R.A.~and Dec~$13^{\rm h}48^{\rm m}56.7^{\rm s}$, $+26^{\rm\circ} 39^{\rm\prime}46.5^{\prime\prime}$. The magnitudes are   $u_s$=22.89 $\pm$ 0.15 (the photometric uncertainty is largely dominated by zero-point uncertainty for the $u_s$-filter),  $g_s$=22.77 $\pm$ 0.01, $r_s$=22.76 $\pm$ 0.02, $i_s$=22.99 $\pm$ 0.07, and $z_s$=22.75 $\pm$ 0.11 from the  GTC/HiPERCAM observations. cNE has a Kron magnitude of  22.82 $\pm$ 0.05 in the FORS2 $R$-filter. A fainter possible host candidate to the west of the FXT (hereafter cW) that lies  at R.A. = $13^{\rm h}48^{\rm m}56.2^{\rm s}$, Dec= $+26^{\rm\circ} 39^{\rm\prime}45.1^{\prime\prime}$ was detected in the  $u_s$, $g_s$, $r_s$, $i_s$, $z_s$-band images of the GTC/HiPERCAM observation. Aperture photometry of cW for a 2.4$^{\prime\prime}$ aperture diameter gives a magnitude of $u_s$=25.6 $\pm$ 0.2, $g_s$=25.08 $\pm$ 0.07, $r_s$=24.71 $\pm$ 0.08, $i_s$=23.9 $\pm$ 0.1, and $z_s$=23.4 $\pm$ 0.1. 

We determined the density of sources brighter than or as bright as the candidate host cX in the HiPERCAM $g_s$ filter to be 0.027 objects/sq.~arcsec in a region of 25\arcsec{} $\times$ 25\arcsec{} centred on R.A. = $13^{\rm h}48^{\rm m}56.5^{\rm s}$, Dec = $+26^{\rm\circ}39^{\rm\prime}44.8^{\prime\prime}$. Assuming Poisson statistics and  considering the 1\arcsec{} error region, we compute the probability of a chance alignment between \xrt{} and cX to be $\sim$7.8\%. Similarly for cNE and cW we computed the chance alignment probability considering the 4.2\arcsec{}  and 3.6\arcsec{} offset from the center of the X-ray position of \xrt, respectively. cNE has a chance alignment probability of  $\sim$8.1\%, whereas it is  $\sim$16\% for cW.

\subsubsection{Limiting magnitude and completeness limit}
We did not find any transient light (counterpart) to the X-ray transient \xrt{} in our optical observations. For the second epoch GTC/HiPERCAM and VLT/FORS2 data,
we determine the 3$\sigma$ completeness and limiting magnitude using the methods explained in \citet{2022MNRAS.514..302E}. The completeness limits for the images taken in the corresponding filters are   $u_s$=24.7, $g_s$=25.7, $r_s$=25.0, $i_s$=23.9, $z_s$=23.8, and $R$=22.4 whereas the limiting magnitudes are $u_s$=26.2, $g_s$=27.0, $r_s$=26.1, $i_s$=24.4, $z_s$=24.7, and $R$=24.7.

\subsection{Spectroscopic redshift of cNE}
\begin{figure*}
\centering
    \includegraphics[scale=0.43]{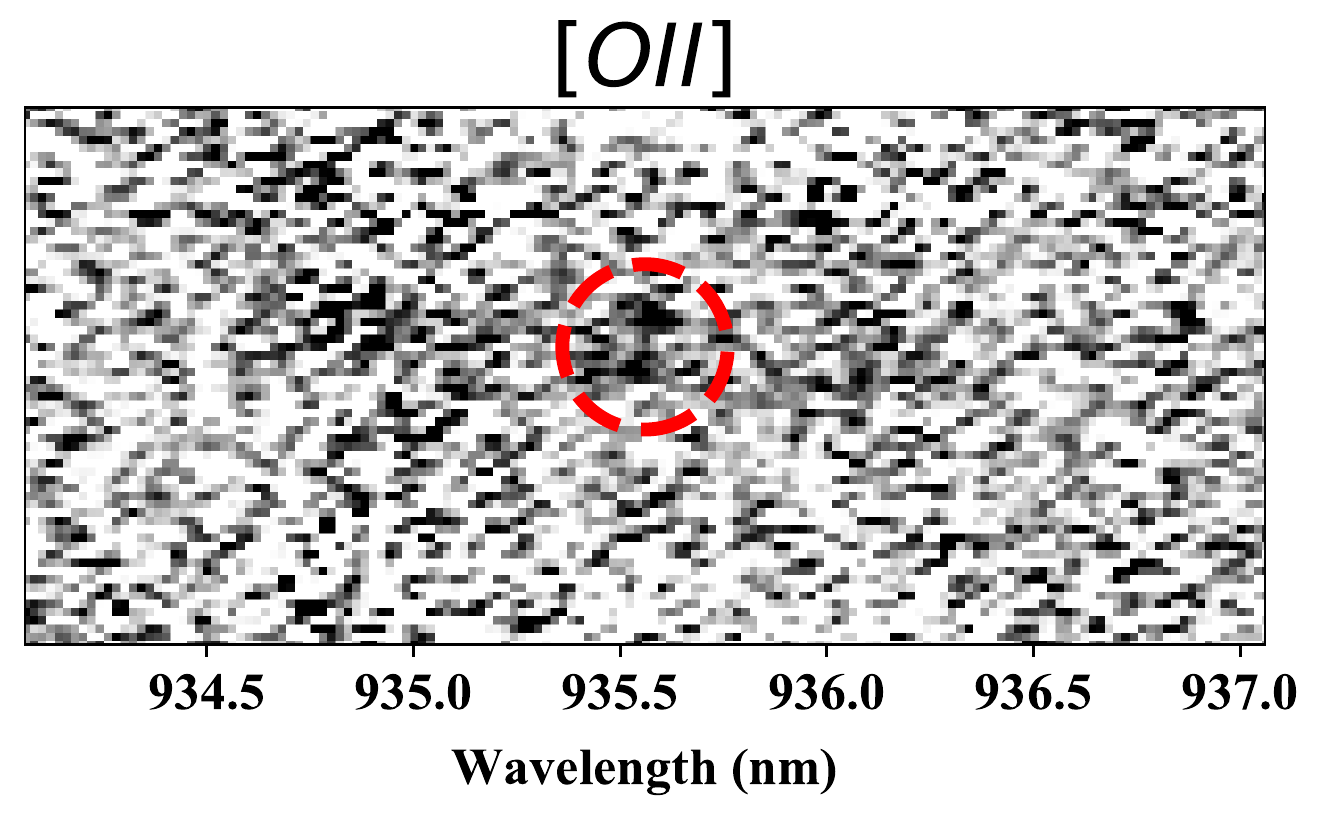}
    \includegraphics[scale=0.43]{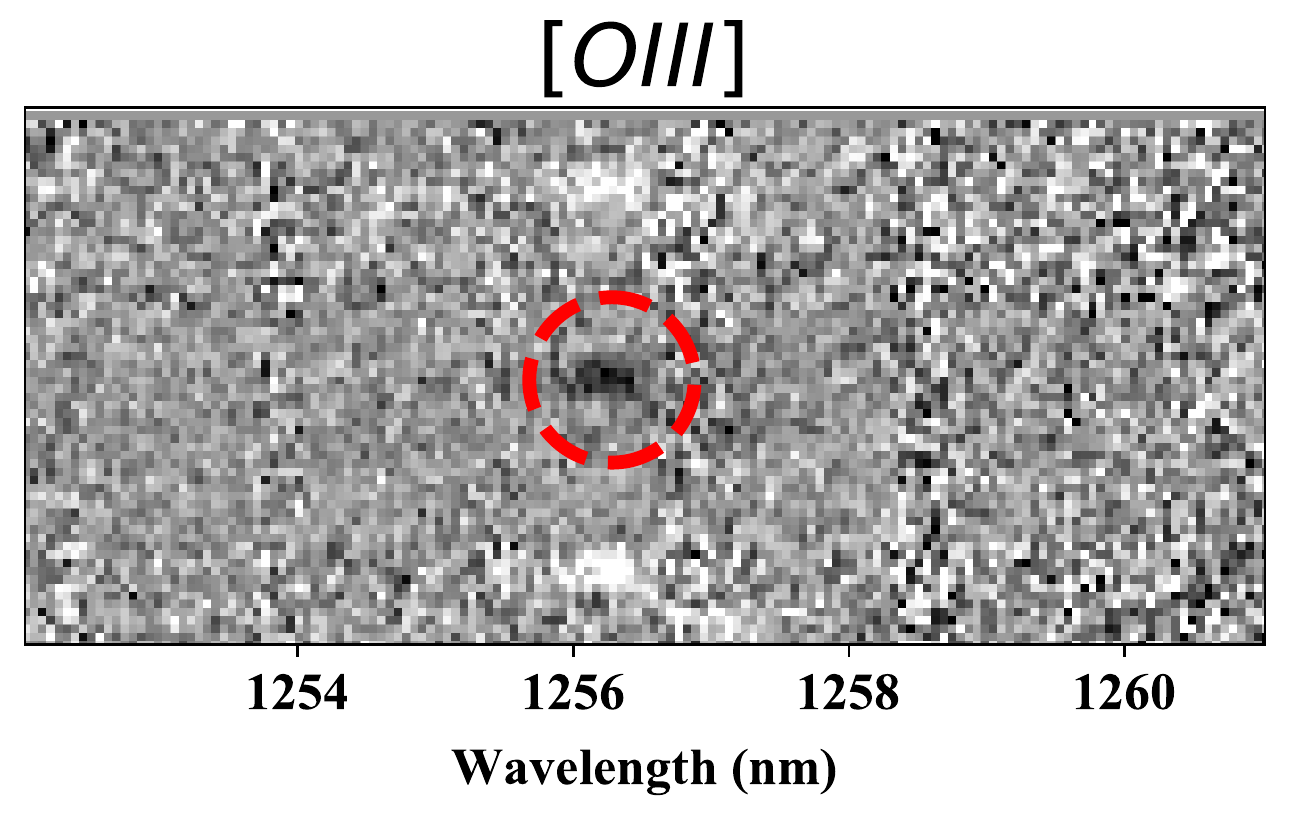}
    \includegraphics[scale=0.43]{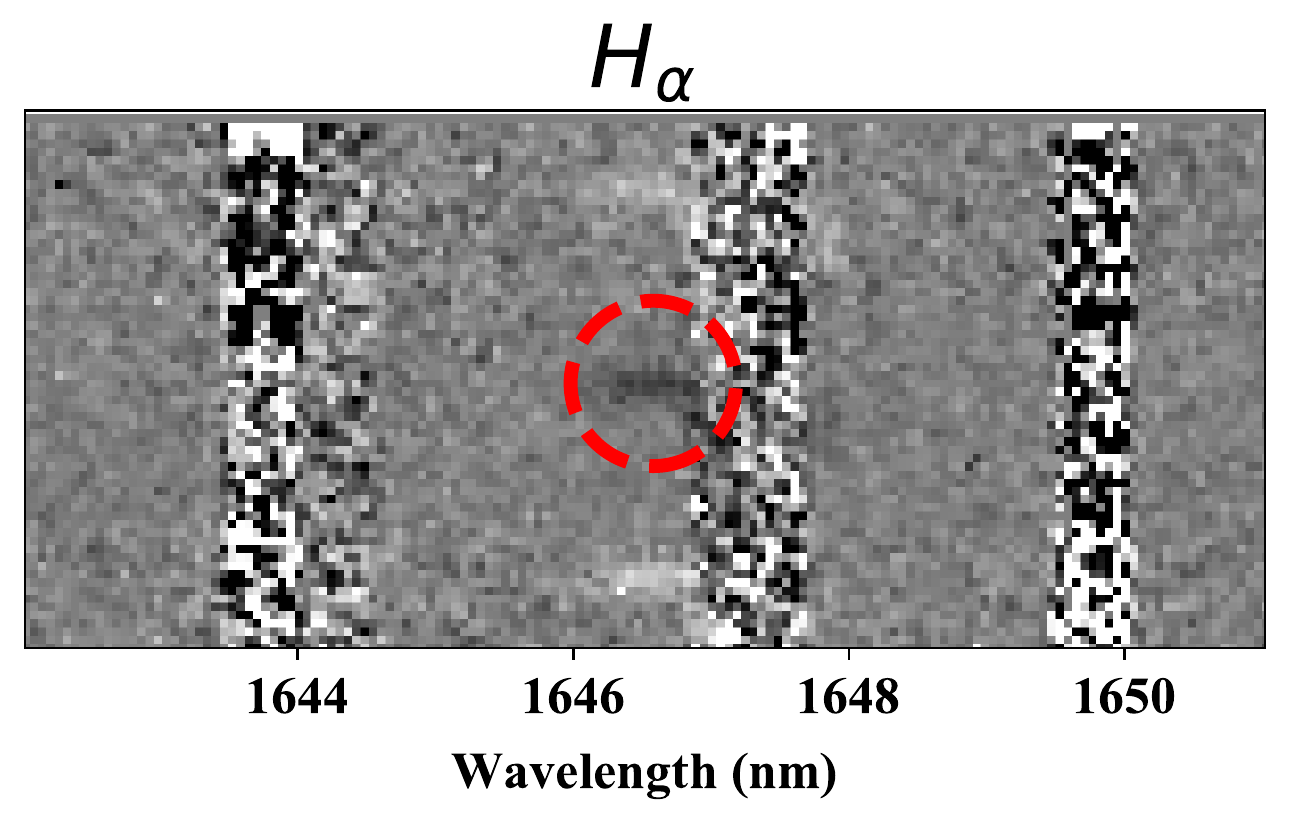}
    \vspace{-0.1cm}
\caption{Three parts of the two dimensional X-shooter spectrum of cNE (SDSS J134856.75+263946.7). We  zoomed-in on the faint emissions lines [OII] doublet with rest wavelengths $\lambda$3727.1$\AA$ and $\lambda$3729.9$\AA$ [OII], [O III] $\lambda$5008.2$\AA$  and H$\alpha$ $\lambda$6564$\AA$  at a redshift $z=1.5082$ marked by red dashed circles. The white horizontal lines seen in the two dimensional spectra are caused by nodding the telescope along the slit, where the Y-axis is the spatial direction along the slit. }
\label{fig:X-shooter}
\end{figure*}

From the X-shooter spectrum of the galaxy cNE (SDSS J134856.75+263946.7), we determine its redshift to be 1.5082 $\pm$ 0.0001, using the faint emissions lines at  $\lambda$16465$\AA$ (restframe wavelength\footnote{https://classic.sdss.org/dr6/algorithms/linestable.html} $\lambda$6564.6$\AA$ H$\alpha$),  $\lambda$12562$\AA$ (restframe wavelength $\lambda$5008.2$\AA$ [O III]; the
$\lambda$4960.3$\AA$ rest-wavelength emission line is not visible in the two dimensional spectra since the $\lambda$4960.3$\AA$ line strength is typically a factor of 1.5 lower compared to the observed $\lambda$5008.2$\AA$ emission line which is just detected) and  $\lambda$9356$\AA$ (restframe wavelength $\lambda$3729.9$\AA$ [OII];  $\lambda$3727.1$\AA$ is overlapped with a sky line). The centre of the emission lines determined by eye from the two-dimensional spectra were taken for the calculation, and the associated uncertainty is derived from the standard deviation on the redshift calculated using the three different emission lines identified. The two dimensional spectra zoomed-in on these faint emission lines are shown in the Figure~\ref{fig:X-shooter}. This redshift is consistent with that reported by \citet{2021ATel14641....1A}  and \citet{2021TNSAN.160....1J}.
The white lines seen in the two dimensional spectra are caused by the AB nodding on slit, where Y-axis is the spatial direction along the 11\arcsec{}--long slit.

\vspace{5 mm}

\subsection{Photometric redshifts and galaxy properties}
To derive the properties of the galaxies cNE and cW, we employed the code \texttt{BAGPIPES} (Bayesian Analysis of Galaxies for Physical Inference and Parameter EStimation; \citealt{2018MNRAS.480.4379C}). Using the \textsc{MultiNest} sampling algorithm, \texttt{BAGPIPES} fits stellar population models taking into account the star formation history and the transmission function of neutral/ionized ISM  for broadband photometry and spectra. Posterior probability distributions for the host galaxy redshift ($z$), age, extinction by dust ($A_V$), star formation rate (SFR), metallicity ($Z$), stellar mass ($M_*$), and specific star formation rate (sSFR) are determined through the fit by \texttt{BAGPIPES}. We used an exponentially declining star formation history function and the parametrization developed by \citet{2000ApJ...533..682C} to account for dust attenuation in the spectral energy distributions (SEDs), fitting for $A_V$ values between 0.0 and 2.0 mag as priors.

\begin{figure*}
    \centering
    \includegraphics[scale=0.43]{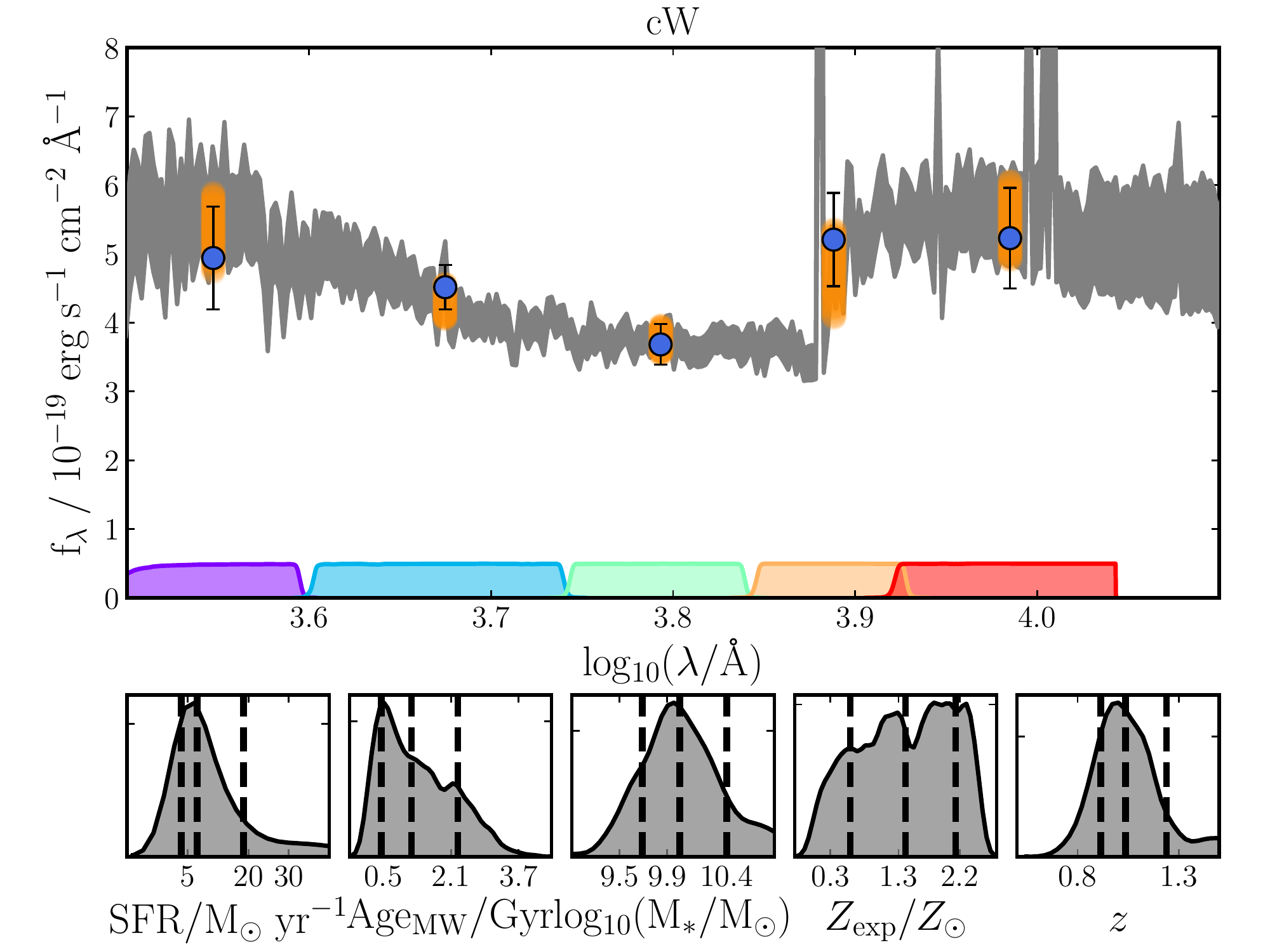}
    \includegraphics[scale=0.37]{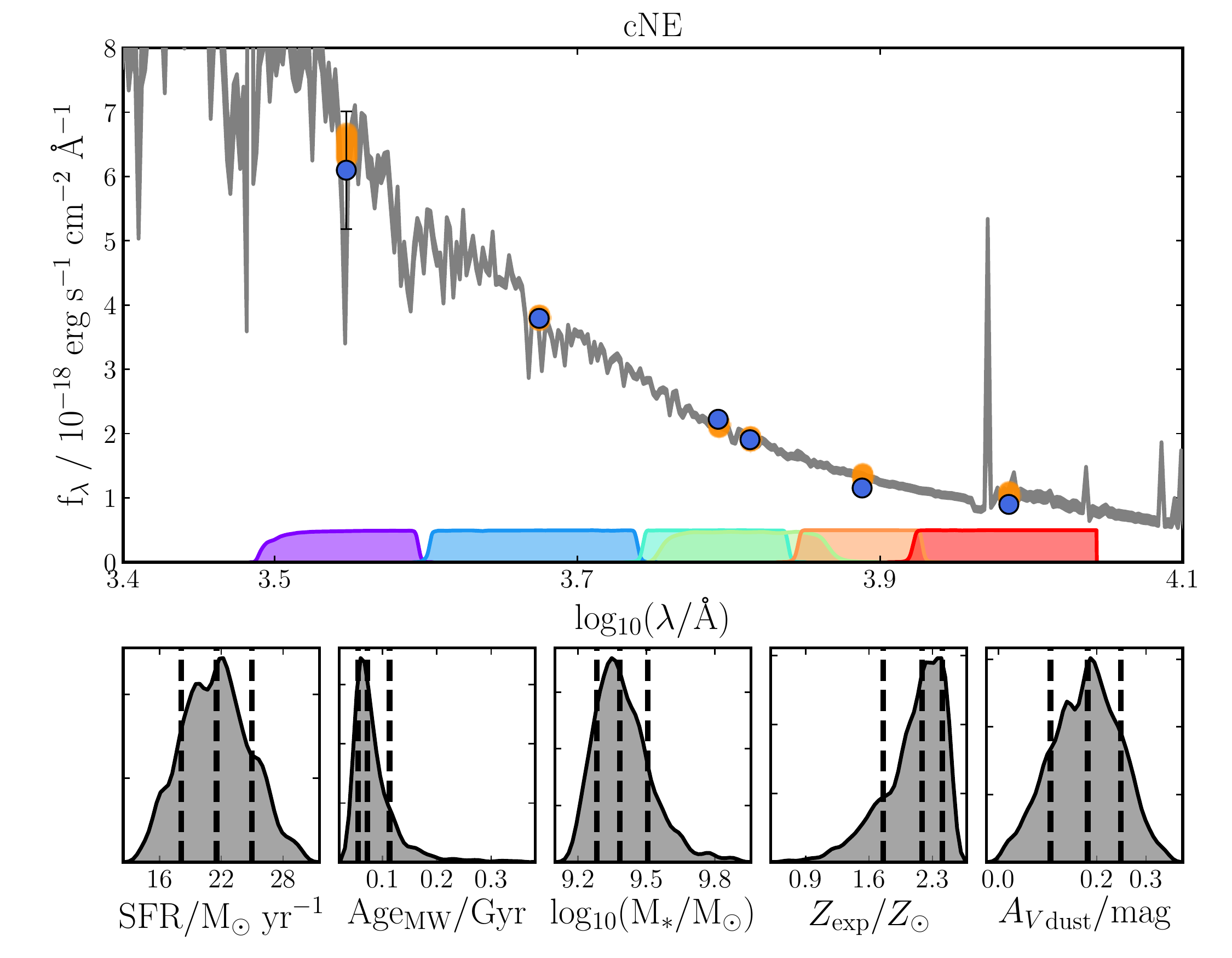}
    \vspace{-0.1cm}
    \caption{Best fitting SED models obtained using the \texttt{BAGPIPES} package. The 16th to 84th percentile range for the posterior probability for the spectrum and photometry (shaded gray and orange, respectively) is shown. The input, observed, photometric data and their 1$\sigma$ uncertainties are given by the blue markers. The wavelength range covered by the photometric filters is marked by the coloured bars at the bottom of the plot. \textit{Bottom panels:} Posterior probability distributions for the five fitted parameters (SFR, age, galaxy stellar mass, metallicity, and, for cW only, redshift). The 16th, 50th, and 84th percentile posterior values are indicated by vertical dashed black lines. The left and right figures shows the best fitting SED models for the galaxies cW and cNE, respectively. 
    }
    \label{fig:SED_models}
\end{figure*}

Figure~\ref{fig:SED_models} shows the 16th to 84th percentile range for the posterior probability  distribution for the spectrum and broadband photometry (shaded \textit{gray} and \textit{orange}). The used input, observed, photometric data are shown in \textit{blue}.  We determine the photometric redshift of cW to be $z_{\rm phot}=1.04^{+0.22}_{-0.1}$. The posterior probability  distribution of the other fitted parameters are shown in the bottom panels of Figure~\ref{fig:SED_models}. Similarly, for the galaxy cNE, we obtained the posterior probability distribution for the fitted parameters, considering a fixed redshift of $z_{spec}$ = 1.5082 (see \S~3.2 and Figure~\ref{fig:SED_models}).

\subsection{Galaxy offsets}

We calculated the galaxy offset for the candidate host galaxies cX, cNE and cW in the $g_s$-band image. The candidate host cX lies at an offset of 0.6$\pm 1^{\prime\prime}$. This corresponds to a physical offset of 0.7 $\pm$ 1.2 kpc, if the candidate host galaxy is at the distance of the Abell 1795 cluster ($z$=0.063). For an assumed redshift between 1 and 1.5, the physical offset ranges from 4.5 to 4.8 kpc. We computed the offset between the galaxy cNE and the center of the localization uncertainty region of \xrt{} to be 4.2$\pm1^{\prime\prime}$. At the redshift of cNE this angular offset translates to a physical projected offset of 37$\pm$9 kpc\footnote{We use https://www.astro.ucla.edu/wright/CosmoCalc.html}.
Similarly, for the galaxy cW, we derive the angular offset to be 3.6$\pm1^{\prime\prime}$. Given the photometric redshift of 1.04 (see \S~3.3), this implies a distance of $\sim$1700 Mpc, which yields a physical offset of 30$\pm$8 kpc. The offsets of the candidate host galaxies cX, cNE and cW are shown in Figure~\ref{fig:210423}.  

\vspace{5mm}
\section{Discussion}\label{sec:discuss}

In this manuscript we report on the X-ray detection of the FXT \xrt. We furthermore discuss our optical observations. No optical counterpart was discovered. We do detect candidate host galaxies and we discuss their properties and implications for the nature of this FXT.
The X-ray light curve of \xrt{} is well fit by a broken power law. The index before the break is --0.17$\pm$0.06 and this part is called the plateau phase (see Figure~\ref{fig:210423_LC_SPEC}).
\xrt{}'s light curve shares many similarities with that of several other FXTs such as \cdf{} (\citealt{Xue2019}),  XRT~100831 (\citealt{Quirola2022}), XRT~170901 (\citealt{2022ApJ...927..211L}), XRT~030511 and  XRT~110919 (\citealt{Quirola2022}; \citealt{2022ApJ...927..211L}). One of the explanations favoured by \citet{Xue2019} for \cdf{} is a magnetar-powered fast X-ray transient in the aftermath of the BNS merger. If true, and given the similarities in the light curve shape, the same explanation could be used for \xrt{}. \citet{2021ApJ...915L..11A} indeed suggest that the \xrt{} data is consistent with the magnetar models of \citet{2013ApJ...763L..22Z}, \citet{2014MNRAS.439.3916M}, \citet{2016ApJ...819...14S}, \citet{2016ApJ...819...15S}, and \citet{2017ApJ...835....7S}. \citet{2017ApJ...835....7S}
defines three zones where the observer would detect a different light curve shape: if the observer's line of sight is in the jet zone, they would see a beamed SGRB signal, while if it is in the free zone, the observer can detect the FXT but is unlikely to detect a $\gamma$- ray burst. In the trapped zone, the X-ray emission first has to ionize the ejecta material but when the ejecta becomes optically thin the X-ray radiation will become detectable to an external observer. Assuming this model is true, the light curve of \xrt{} suggests that our line of sight crosses the free zone.

The power-law decay after the break for \xrt{} is $-$3.8 $\pm$ 1.2, which is consistent within 3$\sigma$ with values found for other FXTs where the light curve can also be well fit with a broken power law (\citealt{Xue2019}; \citealt{Quirola2022}; \citealt{2022ApJ...927..211L}) . For those FXTs, the power law index ranges between $-$1.6 and $-$2.2 (median value of $-$1.9 $\pm$ 0.3). The light curve decay index of \xrt{} and similar FXTs is steeper than that of the median afterglow light curve decay slope for GRBs ($\sim-$1.2; \citealt{2009MNRAS.397.1177E}). \citet{2017ApJ...835....7S}, in their magnetar model suggest that the decay index might become steeper (steeper than $-$3) if the magnetar collapses to a black hole (\citealt{2017ApJ...835....7S}; \citealt{2021ApJ...915L..11A}). If the magnetar does not collapse to a black hole, the X-ray light curve in the free zone would follow a decay with a power law index between -1 and -2 (\citealt{2017ApJ...835....7S}).

A host association and a host redshift determination would help determine the (peak) luminosity, and energy associated with the transient and will allow to convert an angular to a physical offset. Constraints on any or all of these parameters will help constrain the FXT nature. Therefore, we discuss in detail the implications for each of the three possible host galaxies for \xrt{}.

\begin{figure*}
    \centering
    \includegraphics[scale=0.60]{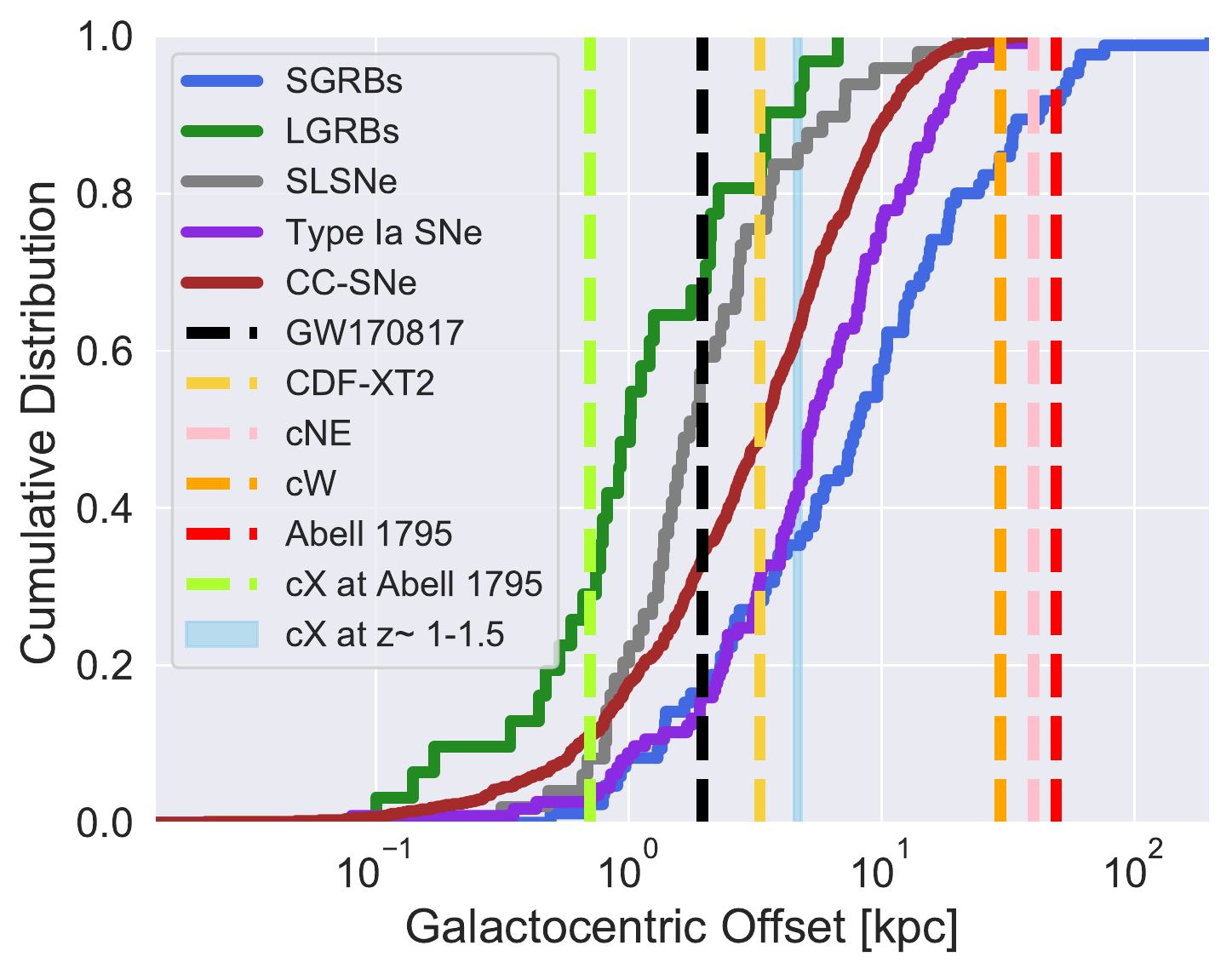}
    \includegraphics[scale=0.56]{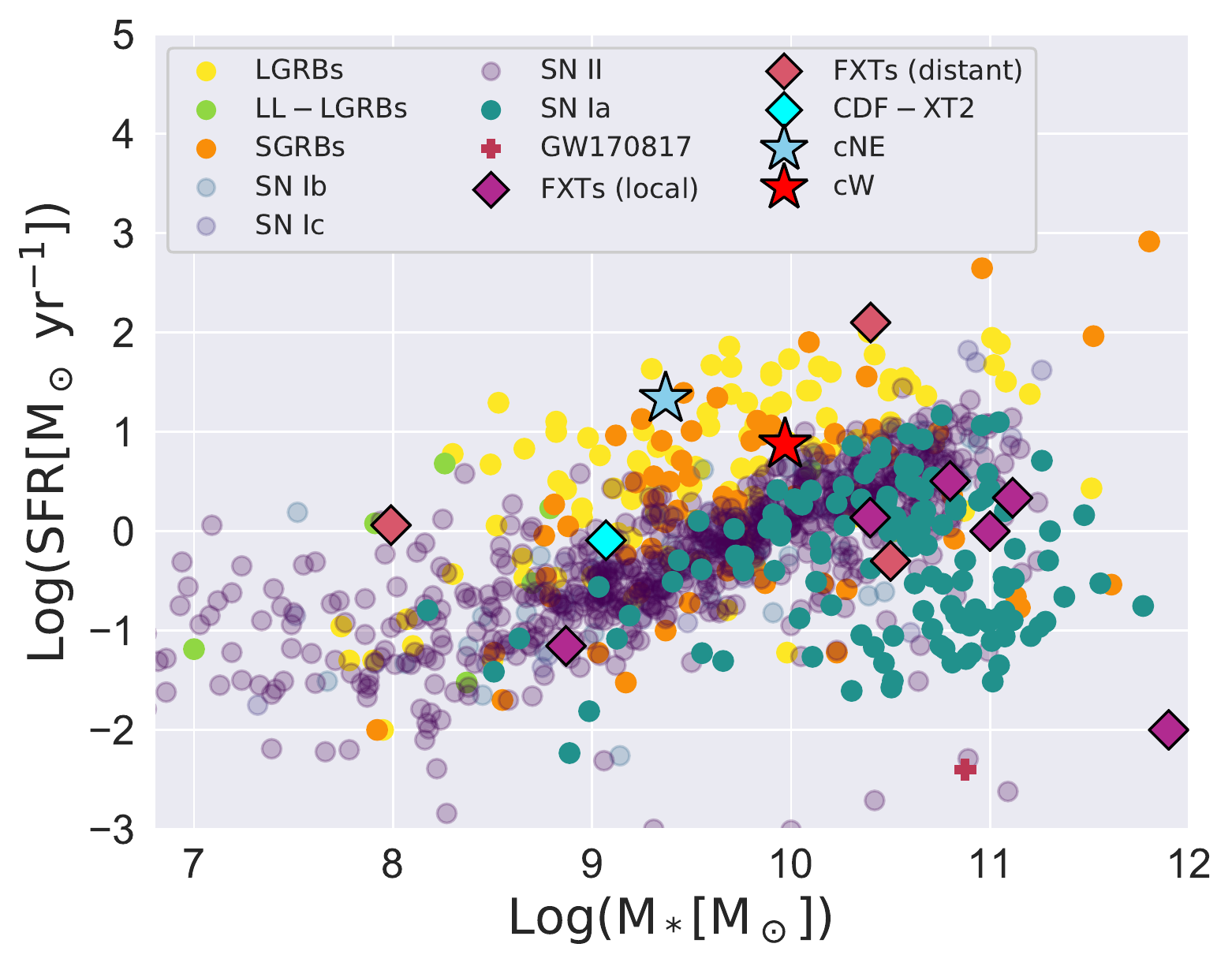}

    \caption{\textit{Left panel:} Projected offset of the FXT \xrt{} with respect to the candidate host galaxies cNE (pink dashed line) and cW (orange dashed line) in comparison with the cumulative distribution of galactocentric offset in kpc seen in SGRBs (blue line; \citealt{2022arXiv220601763F}), long gamma-ray bursts (green line; \citealt{2017MNRAS.467.1795L}), superluminous supernovae  (SLSNe; gray line; \citealt{2021ApJS..255...29S}), Type Ia supernovae (violet line; \citealt{2020ApJ...901..143U}) and core-collapse supernovae (CC-SNe; brown line; \citealt{2012ApJ...759..107K}, \citealt{2021ApJS..255...29S}). The dashed red line indicates the projected offset of \xrt{} with respect to the nearest galaxy LEDA~94646 of the Abell 1795 cluster. The dashed yellow-green and light blue lines shows the galactocentric offset of candidate host galaxy cX if it lies at the distance of Abell 1795 or  if it has an assumed redshift between z $\sim$ 1--1.5, respectively. The projected offsets for GW170817 (\citealt{2017ApJ...848L..28L}) and CDF-XT2 (\citealt{Xue2019}) are shown with the black and yellow dashed lines. \textit{Right panel:}  M$_{*}$ and SFR of the two candidate host galaxies cNE and cW (coloured stars) compared with host galaxies of other transient events. FXT candidates reported by \citet{Quirola2022} and CDF-XT2 (cyan; \citealt{Xue2019}) are shown with diamonds. Coloured circles represent LGRBs, SGRBs, low-luminosity LGRBs (see the references in figure 13 of \citealt{Quirola2022}; \citealt{2022arXiv220601763F}), SN Ia, SN Ib, SN Ic and SN II (\citealt{2014A&A...572A..38G}, \citealt{2021ApJS..255...29S}). GW\,170817 is marked by the red \textbf{+} sign (\citealt{Im2017}).} 

    \vspace{-0.1cm}
    \label{fig:sGRB}
\end{figure*}

\subsection{ Assuming cX is the host galaxy for \xrt{}}
\label{cX}
In our second epoch GTC/HiPERCAM observations, we detect the candidate host galaxy cX with the $g_s$ = 25.9$\pm$0.1 within the $\sim$1\arcsec{} X-ray uncertainty region (see Section~\ref{sec:results_optical}). The lack of detections at other photometric bands preclude the determination of the photometric redshift. If at the distance of the Abell 1795 cluster ($z$=0.063), cX lies at a (projected) offset of 0.7 $\pm$ 1.2 kpc from the FXT. Alternatively, if cX is at the distance of the galaxies cNE or cW, the (projected) offset would be 4.8 $\pm$ 8.7 kpc  or 4.5 $\pm$ 8.2 kpc, respectively. The \textit{left} panel of Figure \ref{fig:sGRB} shows the cumulative host galaxy offset distribution of SGRBs (\citealt{2022arXiv220601763F}), long GRBs (\citealt{2017MNRAS.467.1795L}), superluminous supernovae (\citealt{2021ApJS..255...29S}), Type Ia SNe (\citealt{2020ApJ...901..143U}) and core-collapsed SNe (CC-SNe;  \citealt{2012ApJ...759..107K}, \citealt{2021ApJS..255...29S}) compared with the offsets (in kpc) of GW170817 (\citealt{2017ApJ...848L..28L}), \cdf{} (\citealt{Xue2019}), and \xrt{}. The possible offsets between \xrt{} and cX are similar to the offsets seen for \cdf{}, SGRBs, LGRBs, SLSNe and Type Ia SNe (see \textit{Left} panel of Figure~\ref{fig:sGRB}). 

If cX is at the distance of the Abell cluster, it has an absolute magnitude of $M_g$ = -11.4. Given this absolute magnitude and the fact that cX seems an extended source, we can rule out a globular cluster host considering the observed  absolute magnitudes and sizes of globular clusters (\citealt{2019ARA&A..57..375S}). We cannot rule out the possibility that the candidate host galaxy is a faint dwarf galaxy at the distance of the Abell cluster. If so, then the X-ray flare could be due to an IMBH-WD TDE (\citealt{2020SSRv..216...39M}). If cX lies at the distance of cNE or cW, the candidate host galaxy would have an absolute magnitude of -19.3 or -18.2, respectively. We cannot discard a dwarf galaxy association even if at such a redshift as these absolute magnitudes are similar to that of the Large Magellanic Cloud (\citealt{2022NatAs...6..911B}).

\begin{figure*}
    \centering
    \includegraphics[scale=0.72]{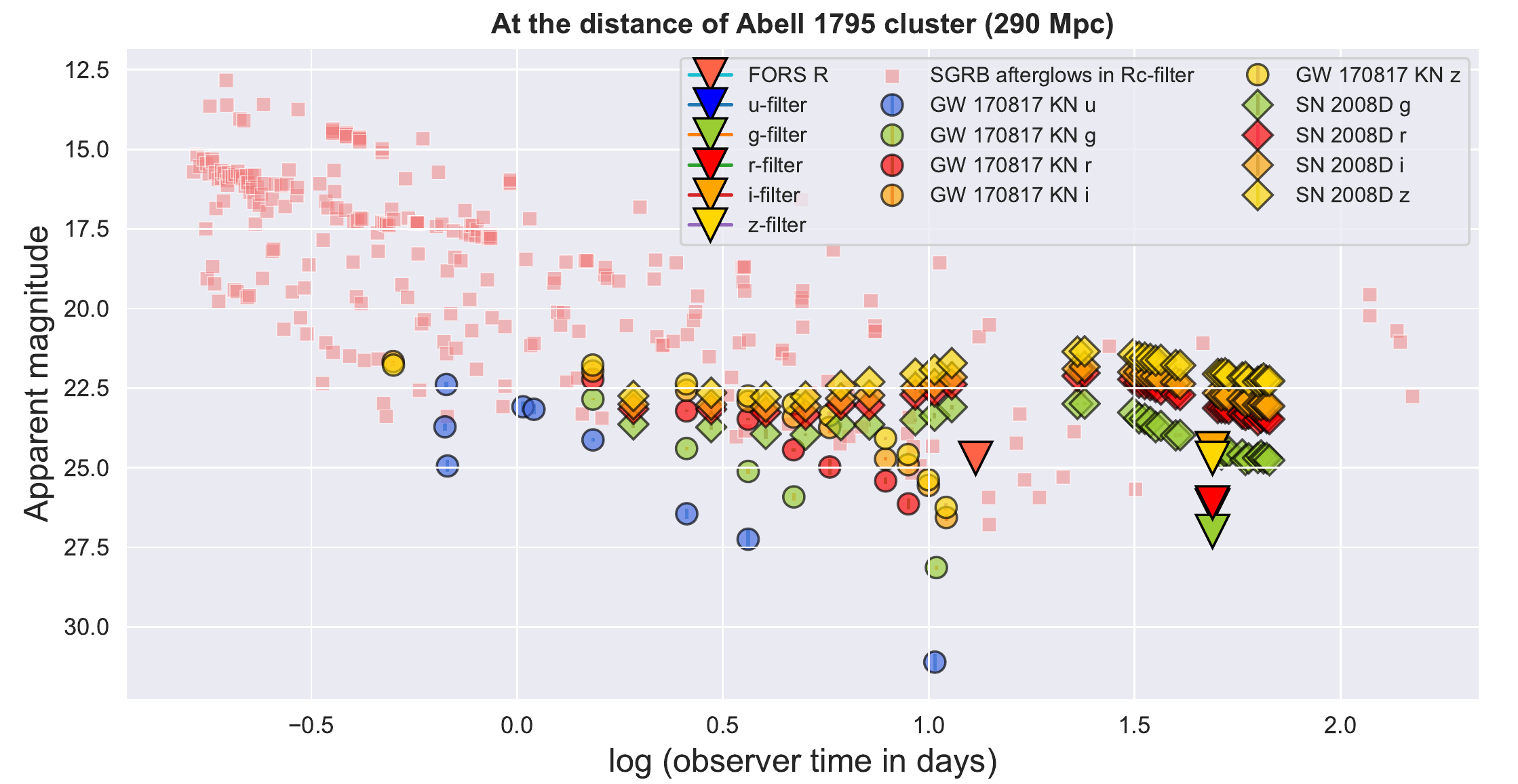}
    \vspace{0.1cm}
    \centering
    \includegraphics[scale=0.72]{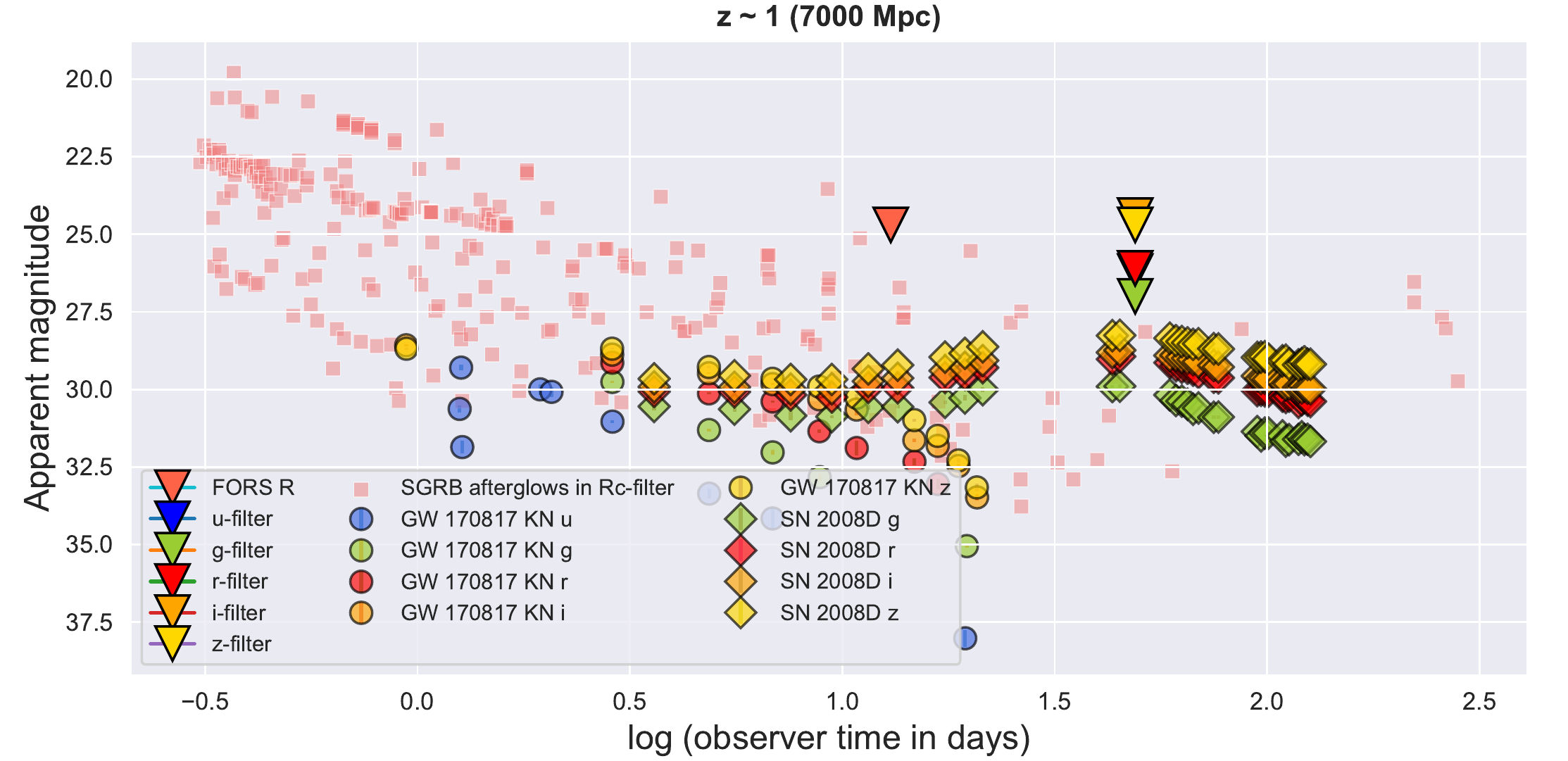}
    \caption{We compare our deep optical limits (shown with coloured triangles; the symbol for the $u$-filter falls behind that of the $r$-filter and is thus not visible) derived from the follow-up observations of \xrt{} using VLT/FORS2 and GTC/HiPERCAM data 13 and 55 days after the event, respectively, with the light curve of the kilonova associated with GW~170817 observed in different filters (GW~170817 KN; \citealt{2017ApJ...848L..17C}) as shown with differently coloured circles. The different filter light curves of the supernova associated with the SBO, SN~2008D (\citealt{2008Natur.453..469S}), are shown using diamond symbols. Again, observations in different filters are marked using different colours. We plotted both light curves for two distances, where in the \textit{top panel} we assume the distance of Abell~1795, and in the  \textit{bottom panel} we assume the distance associated with a redshift, $z=$1. Considering the supernova signal following a SN SBO, such as SN~2008D, it is unlikely that we did not detect a counterpart if \xrt{} occurred at the distance of the Abell~1795 cluster. Our VLT/FORS R-filter observation is deeper than the apparent magnitude the supernova SN~2008D would have had if it occurred at the distance of the Abell~1795 cluster. However, in order to detect a kilonova such as that in GW~170817, we need lower latency and deeper observations if a kilonova is associated with the FXT \xrt{}. To detect either an SBO or kilonova at the $z\sim$1, we need deeper and more prompt observations within 10 days. } 

    \vspace{-0.1cm}
    \label{fig:optical limits}
\end{figure*}

 \vspace{8mm}

\subsection{Assuming cNE is the host galaxy for \xrt}
If \xrt{} is associated with the candidate host galaxy cNE, which has a luminosity distance of $\sim$11 Gpc, the transient would have a peak X-ray luminosity of $\sim$ $10^{46}$ \lum. This is consistent with the suggestion of \citet{2014ARA&A..52...43B} that BNS mergers have an X-ray peak luminosity of $L_{\rm X,peak}{\approx}$10$^{44}$--10$^{51}$~erg~s$^{-1}$ (considering beamed emission for larger luminosity). The \xrt{} luminosity at this distance is also consistent with the luminosity of \cdf{} for which a distance has been determined through a host galaxy association (\citealt{Xue2019}) .  
The offset of \xrt{} with respect to the candidate host galaxy cNE is consistent with the 90th percentile of the SGRB offsets. The parameters such as SFR and stellar mass of cNE, obtained through the SED fitting using \texttt{BAGPIPES} are shown in Figure~\ref{fig:sGRB}. In the \textit{right panel} of Figure~\ref{fig:sGRB} we show the stellar mass (M$_{*}$) and star-formation rate (SFR) of the candidate host galaxies cNE (coloured star) compared with host galaxies of other transient events. The FXT sample within 100 Mpc (\citealt{Quirola2022}) tends to have more massive hosts and  lower SFR rates than more distant FXTs.
The values for the XRT~041230, XRT~080819 and  XRT~141001 for the sample of FXTs with distances larger than 100 Mpc have a larger spread than that for the sample of FXTs within 100 Mpc. cNE has a high SFRs compared to the candidate hosts of most FXTs.

The star formation rate and the mass of cNE are  consistent with some of the hosts of SGRBs and LGRBs (orange circles and yellow circles, respectively, in the \textit{right panel} of Figure~\ref{fig:sGRB}). An SBO origin of the \xrt{} is very unlikely given the low probability to have such a high peak luminosity ($L_{\rm{X,peak}} \approxlt 10^{45}$ \lum{}  for supernova SBOs; \citealt{2008Natur.453..469S}; \citealt{2017hsn..book..967W}; \citealt{2022ApJ...933..164G}), offset (see Figure~\ref{fig:sGRB}; \citealt{2020ApJ...901..143U}; \citealt{2012ApJ...759..107K}; \citealt{2021ApJS..255...29S}), $M_*$ and SFR (see Figure~\ref{fig:sGRB}; \citealt{2014A&A...572A..38G}).

\subsection{Assuming cW is the host galaxy}
The photometric redshift of cW $z_{phot}\approx 1.04^{+0.22}_{-0.14}$. 
If \xrt{} lies at the distance of cW, it had a peak luminosity of (3--7) $\times$ 10$^{45}$ \lum~and a projected  physical offset of 30$\pm$ 8 kpc (pink line in the \textit{Left} panel of Figure~\ref{fig:sGRB}). A few SGRBs (orange circles in the \textit{right panel} of Figure~\ref{fig:sGRB}) originate from hosts with similar SFR  and $M_*$  (cW has a  SFR $\sim$ 7.5 $\msun{}$ yr$^{-1}$ and $M_*$ [$\msun{}$] $\sim$ 10$^{10}$; \textit{Right} panel of Figure \ref{fig:sGRB}). As in the case of cNE, the host galaxy SFR and M$_{\star}$, the offset between the center of the host and the FXT uncertainty region, and the  peak X-ray luminosity agree with the values found for SGRBs making a BNS merger scenario possible for \xrt{}. We discard an SBO origin for \xrt{} based on the high X-ray luminosity and the large physical offset.

\subsection{Assuming \xrt{} is associated with Abell~1795}

We already briefly discussed a possible association of \xrt{} with the galaxy cluster Abell~1795 in Section~\ref{cX}. However, even if \xrt{} is not associated with cX it can be associated with Abell~1795, which is at $z$=0.063 ($\sim$290 Mpc). This is what we discuss here.
If \xrt{} originated at the distance of the Abell~1795 cluster, the projected separation to the closest catalogued cluster galaxy, LEDA 94646 (R.A.:$13^{\rm h}48^{\rm m}53.6^{\rm s}$, Dec: $+26^{\rm\circ} 39^{\rm\prime}52^{\prime\prime}$) is $\sim$49.1 kpc (\textit{bottom right} panel of Figure~\ref{fig:210423}). Although this large offset is rare, but not impossible for an SGRB host. On the other hand, the associated offset is too large for a core-collapse SN or a type Ia SN origin of \xrt{}. For the distance of 290 Mpc, \xrt{} had a peak luminosity of 7$\times$ 10$^{42}$ \lum. We cannot rule out any of the progenitor models we consider here considering the peak X-ray luminosity. But, if associated with a typical type Ia supernova (peak absolute magnitude of -19 in V-band; \citealt{2000ARA&A..38..191H} or a CC-SNe, peak absolute magnitude of -17.5 in B-band; \citealt{2014AJ....147..118R})  at the distance of the cluster, the optical supernova would have a peak magnitude of $m_V$ $\sim$ 18.5 and $m_B$ $\sim$ 20, respectively. Therefore, it is unlikely that the associated optical supernova  went undetected in our deep observations (see Figure \ref{fig:optical limits}). 
\section{Conclusion}\label{sec:conclusion}
We present the X-ray light curve and spectrum and the search for a possible host galaxy of the FXT \xrt. We consider various candidate hosts for \xrt, namely: a candidate host galaxy, called cX, detected in the $g_s$-filter  of the GTC/HiPERCAM image with a magnitude of 25.9 $\pm$ 0.1 that falls within the 3$\sigma$ X-ray uncertainty region, a candidate host at a redshift, $z_{spec}$ $\approx$ $1.51$, named cNE, and a candidate host at a redshift,  $z_{phot}$ $\approx$ $1.04$, named cW. Finally, we also investigate if  \xrt{} can be associated with the galaxy cluster Abell~1795. Considering the  association of \xrt{} with the candidate host cX in the 3-sigma X-ray uncertainty region,  both a BNS and a WD-IMBH TDE origin remain plausible. For all host scenarios, we can rule out that the FXT originates in an SN SBO. The galactocentric offset, star formation rate, mass, and luminosity of the candidate hosts cW and cNE are consistent with a BNS merger origin. The plateau in the X-ray light curve is similar to that of the FXT \cdf{}, also seen in some SGRBs, consistent with a BNS merger origin.

\section{Acknowledgements}
DE acknowledges discussions with Ashley Chrimes, Fiorenzo Stoppa, Nicola Gaspari and Anne Inkenhaag. FEB acknowledges support from ANID-Chile BASAL CATA ACE210002 and FB210003, FONDECYT Regular 1200495 and 1190818, and Millennium Science Initiative Program  – ICN12\_009. MF is supported by a Royal Society - Science Foundation Ireland University Research Fellowship. JQV acknowledges support from ANID grants Programa de Capital Humano Avanzado folio \#21180886, CATA-Basal AFB-170002, and Millennium Science Initiative ICN12\_009. Based on observations collected at the European Southern Observatory under ESO programme 105.209R.001 (PI Jonker). We thank the anonymous referee for the helpful comments on this manuscript.

The design and construction of HiPERCAM was funded by the European Research Council under the European Union’s Seventh Framework Programme (FP/2007-2013) under ERC-2013-ADG Grant Agreement no. 340040 (HiPERCAM). VSD and HiPERCAM operations are supported by STFC grant ST/V000853/1. This work is based on observations made with the Gran Telescopio Canarias (GTC), installed at the Spanish Observatorio del Roque de los Muchachos of the Instituto de Astrofísica de Canarias, in the island of La Palma. The Pan-STARRS1 Surveys (PS1) and the PS1 public science archive have been made possible through contributions by the Institute for Astronomy, the University of Hawaii, the Pan-STARRS Project Office, the Max-Planck Society and its participating institutes, the Max Planck Institute for Astronomy, Heidelberg and the Max Planck Institute for Extraterrestrial Physics, Garching, The Johns Hopkins University, Durham University, the University of Edinburgh, the Queen's University Belfast, the Harvard-Smithsonian Center for Astrophysics, the Las Cumbres Observatory Global Telescope Network Incorporated, the National Central University of Taiwan, the Space Telescope Science Institute, the National Aeronautics and Space Administration under Grant No. NNX08AR22G issued through the Planetary Science Division of the NASA Science Mission Directorate, the National Science Foundation Grant No. AST-1238877, the University of Maryland, Eotvos Lorand University (ELTE), the Los Alamos National Laboratory, and the Gordon and Betty Moore Foundation.

SDSS-IV acknowledges support and 
resources from the Center for High 
Performance Computing  at the 
University of Utah. The SDSS 
website is www.sdss.org. SDSS-IV is managed by the 
Astrophysical Research Consortium 
for the Participating Institutions 
of the SDSS Collaboration including 
the Brazilian Participation Group, 
the Carnegie Institution for Science, 
Carnegie Mellon University, Center for 
Astrophysics | Harvard \& 
Smithsonian, the Chilean Participation 
Group, the French Participation Group, 
Instituto de Astrof\'isica de 
Canarias, The Johns Hopkins 
University, Kavli Institute for the 
Physics and Mathematics of the 
Universe (IPMU) / University of 
Tokyo, the Korean Participation Group, 
Lawrence Berkeley National Laboratory, 
Leibniz Institut f\"ur Astrophysik 
Potsdam (AIP),  Max-Planck-Institut 
f\"ur Astronomie (MPIA Heidelberg), 
Max-Planck-Institut f\"ur 
Astrophysik (MPA Garching), 
Max-Planck-Institut f\"ur 
Extraterrestrische Physik (MPE), 
National Astronomical Observatories of 
China, New Mexico State University, 
New York University, University of 
Notre Dame, Observat\'ario 
Nacional / MCTI, The Ohio State 
University, Pennsylvania State 
University, Shanghai 
Astronomical Observatory, United 
Kingdom Participation Group, 
Universidad Nacional Aut\'onoma 
de M\'exico, University of Arizona, 
University of Colorado Boulder, 
University of Oxford, University of 
Portsmouth, University of Utah, 
University of Virginia, University 
of Washington, University of 
Wisconsin, Vanderbilt University, 
and Yale University.


%









\bibliography{sample631}{}
\bibliographystyle{aasjournal}



\end{document}